\documentclass[longauth]{aa}

\usepackage{graphicx}
\usepackage{txfonts}
\usepackage{amsmath}
\usepackage{color} %costum addition 
\usepackage{xcolor} %costum addition 
\usepackage{makecell} % custom addition
\usepackage[para]{threeparttable} % custom addition

\usepackage{hyperref}
\newcommand*\pyorbit{\texttt{PyORBIT}}
\newcommand{\thirstee}{\texttt{THIRSTEE}}
\newcommand{\whfast}{\texttt{WHFAST}}
\newcommand{\rebound}{\texttt{REBOUND}}

\newcommand*\tess{{\it TESS}}

\newcommand{\snr} {\mbox{S/N}}

\newcommand{\prot}{\mbox{P$_{\rm rot}$}}

\newcommand{\teff}{$T_{{\rm eff}}$}
\newcommand{\kms}{\mbox{km\,s$^{-1}$}}
\newcommand{\ms}{\mbox{m s$^{-1}$}}
\newcommand{\gcm}{\mbox{g cm$^{-3}$}}

\newcommand{\logg} {\mbox{log\,{\it g}}}

\newcommand{\mplanet}{\mbox{$M_{\rm p}$}}
\newcommand{\rplanet}{\mbox{$R_{\rm p}$}}
\newcommand{\rhoplanet}{\mbox{$\rho_{\rm p}$}}

\newcommand{\teq}{$T_{{\rm eq}}$}

\newcommand{\mearth}{\mbox{M$_\oplus$}}
\newcommand{\rearth}{\mbox{R$_\oplus$}}

\newcommand{\msun}{\mbox{M$_\odot$}}
\newcommand{\rsun}{\mbox{R$_\odot$}}
\newcommand{\mstar}{\mbox{$M_\star$}}
\newcommand{\rstar}{\mbox{$R_\star$}}

\newcommand{\rhostar}{\mbox{$\rho_\star$}}
\newcommand{\rhosun}{\mbox{$\rho_\odot$}}

\newcommand{\logRHK}{\mbox{$\log {\rm R}^{\prime}_{\rm HK}$}}

\newcommand{\halpha}{\mbox{H$\alpha$}}

\newcommand{\starname}{\mbox{TOI-771}}

\begin{document}

   \title{A transiting rocky super-Earth and a non-transiting sub-Neptune orbiting the M dwarf TOI-771}

   \author{G. Lacedelli\inst{\ref{iac}}\,$^{\href{https://orcid.org/0000-0002-4197-7374}{\protect\includegraphics[height=0.19cm]{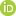}}}$,
          E. Pallé\inst{\ref{iac}, \ref{ull}}\,$^{\href{https://orcid.org/0000-0003-0987-1593}{\protect\includegraphics[height=0.19cm]{orcid.jpg}}}$,
          Y. T. Davis\inst{\ref{birmingham}}\,$^{\href{https://orcid.org/0009-0000-6625-137X}{\protect\includegraphics[height=0.19cm]{orcid.jpg}}}$, 
          R. Luque\inst{\ref{chicago}\fnmsep\thanks{NHFP Sagan Fellow}}\,$^{\href{https://orcid.org/0000-0002-4671-2957}{\protect\includegraphics[height=0.19cm]{orcid.jpg}}}$,
          G. Morello\inst{\ref{iaa}, \ref{inaf_palermo}}, 
          H. M. Tabernero\inst{\ref{ieec},\ref{ice_csic}},
          M. R. Zapatero Osorio\inst{\ref{CSIC-INTA}}\,$^{\href{https://orcid.org/0000-0001-5664-2852}{\protect\includegraphics[height=0.19cm]{orcid.jpg}}}$,
          F. J. Pozuelos\inst{\ref{iaa}},
          D. Jankowski\inst{\ref{poland}},
          G. Nowak\inst{\ref{poland}}\,$^{\href{https://orcid.org/0000-0002-7031-7754}{\protect\includegraphics[height=0.19cm]{orcid.jpg}}}$,
          F. Murgas\inst{\ref{iac},\ref{ull}}\,$^{\href{https://orcid.org/0000-0001-9087-1245}{\protect\includegraphics[height=0.19cm]{orcid.jpg}}}$,
          J.~Orell-Miquel\inst{\ref{texas}, \ref{iac}, \ref{ull}}\,$^{\href{https://orcid.org/0000-0003-2066-8959}{\protect\includegraphics[height=0.19cm]{orcid.jpg}}}$,
          J. M. Akana Murphy\inst{\ref{california}\fnmsep\thanks{NSF Graduate Research Fellow}}\,$^{\href{https://orcid.org/0000-0001-8898-8284}{\protect\includegraphics[height=0.19cm]{orcid.jpg}}}$,
          K. Barkaoui\inst{\ref{liege}, \ref{MIT}, \ref{iac}}\,$^{\href{https://orcid.org/0000-0003-1464-9276}{\protect\includegraphics[height=0.19cm]{orcid.jpg}}}$,
          D. Charbonneau\inst{\ref{harvard}}$^{\href{https://orcid.org/0000-0002-9003-484X}{\protect\includegraphics[height=0.19cm]{orcid.jpg}}}$,
          G. Dransfield\inst{\ref{oxford}, \ref{magdalen},\ref{birmingham}},
          E. Ducrot\inst{\ref{ETH}},
          S.~Gerald\'ia-Gonz\'alez\inst{\ref{iac},\ref{ull}},
          J. Irwin\inst{\ref{cambridge}},
          E. Jehin\inst{\ref{liege}},
          H. L. M. Osborne\inst{\ref{ucl}},
          P. P. Pedersen\inst{\ref{cavendish},\ref{ETH}},
          B. V. Rackham\inst{\ref{MIT}}\,$^{\href{https://orcid.org/0000-0002-3627-1676}{\protect\includegraphics[height=0.19cm]{orcid.jpg}}}$,
          M. G. Scott\inst{\ref{birmingham}}\,$^{\href{https://orcid.org/0009-0006-3846-4558}{\protect\includegraphics[height=0.19cm]{orcid.jpg}}}$,
          M. Timmermans\inst{\ref{birmingham}, \ref{liege}},
          A. Triaud\inst{\ref{birmingham}},
          V. Van Eylen\inst{\ref{ucl}}\,$^{\href{https://orcid.org/0000-0001-5542-8870}{\protect\includegraphics[height=0.19cm]{orcid.jpg}}}$
          }
          
   \institute{Instituto de Astrof\'{i}sica de Canarias (IAC), 38205 La Laguna, Tenerife, Spain \label{iac} 
              \email{glacedelli@iac.es}
    \and Departamento de Astrof\'isica, Universidad de La Laguna (ULL), E-38206 La Laguna, Tenerife, Spain \label{ull} 
    \and School of Physics \& Astronomy, University of Birmingham, Edgbaston, Birmingham B15 2TT, United Kingdom\label{birmingham}
    \and Department of Astronomy and Astrophysics, University of Chicago, Chicago, IL 60637, USA \label{chicago} 
    \and Instituto de Astrof{\'i}sica de Andaluc{\'i}a (IAA-CSIC), Glorieta de la Astronom{\'i}a s/n, E-18008 Granada, Spain\label{iaa}
    \and INAF -- Osservatorio Astronomico di Palermo, Piazza del Parlamento, 1, 90134 Palermo, Italy\label{inaf_palermo}
    \and Institut d'Estudis Espacials de Catalunya (IEEC), Edifici RDIT, Campus UPC, 08860 Castelldefels (Barcelona), Spain \label{ieec}
    \and Institut de Ciències de l'Espai (ICE, CSIC), Campus UAB, c/ de Can Magrans s/n, 08193 Cerdanyola del Vallès, Barcelona, Spain \label{ice_csic}
    \and Centro de Astrobiología (CSIC-INTA), Crta. Ajalvir km 4, E-28850 Torrejón de Ardoz, Madrid, Spain \label{CSIC-INTA}
    \and Institute of Astronomy, Faculty of Physics, Astronomy and Informatics, Nicolaus Copernicus University, Grudzi\c{a}dzka 5, 87-100 Toru\'{n}, Poland \label{poland}
    \and University of Texas at Austin, Department of Astronomy, 2515 Speedway C1400, Austin, TX 78712, USA\label{texas}
    \and Department of Astronomy and Astrophysics, University of California, Santa Cruz, CA 95064, USA \label{california} 
    \and Astrobiology Research Unit, Universit\'e de Li\`ege, All\'ee du 6 Ao\^ut 19C, B-4000 Li\`ege, Belgium\label{liege}
    \and Department of Earth, Atmospheric and Planetary Science, Massachusetts Institute of Technology, 77 Massachusetts Avenue, Cambridge, MA 02139, USA\label{MIT},
    \and Center for Astrophysics | Harvard \& Smithsonian, 60 Garden St., Cambridge, MA 02138, USA\label{harvard}
    \and Department of Astrophysics, University of Oxford, Denys Wilkinson Building, Keble Road, Oxford OX1 3RH, UK\label{oxford}
    \and Magdalen College, University of Oxford, Oxford OX1 4AU, UK\label{magdalen}
    \and Institute for Particle Physics and Astrophysics , ETH Z\"urich, Wolfgang-Pauli-Strasse 2, 8093 Z\"urich, Switzerland\label{ETH}
    \and Institute for Astronomy, University of Cambridge, Madingely Road, Cambridge, CB3 0HA, UK\label{cambridge}
    \and Mullard Space Science Laboratory, University College London, Surrey, UK \label{ucl}
    \and Cavendish Laboratory, JJ Thomson Avenue, Cambridge CB3 0HE, UK\label{cavendish}
             }

   \date{Received 18 February 2025; accepted 24 April 2025}

% \abstract{}{}{}{}{} 
% 5 {} token are mandatory
 
  \abstract
   {The origin and evolution of the sub-Neptune population is a highly debated topic in the exoplanet community. With the advent of \textit{JWST}, atmospheric studies can now put unprecedented constraints on the internal composition of this population. In this context, the \thirstee\ project aims to investigate the population properties of sub-Neptunes with a comprehensive and demographic approach, providing a homogeneous sample of precisely characterised sub-Neptunes across stellar spectral types.}
   {We present here the precise characterisation of the planetary system orbiting one of the \thirstee\ M-dwarf targets, TOI-771 ($ d = 25$~pc, $V = 14.9$~mag), known to host one planet, TOI-771 b, which has been statistically validated using \textit{TESS} observations.}
   {We use \tess, SPECULOOS, TRAPPIST, and M-Earth photometry together with $31$ high-precision ESPRESSO radial velocities to derive the orbital parameters and investigate the internal composition of TOI-771 b, as well as exploring the presence of additional companions in the system.}
   {We derived the precise mass and radius for TOI-771 b, a super-Earth with $R_{\rm b} = 1.36 \pm 0.10 $~\rearth\ and $M_{\rm b} = 2.47_{-0.31}^{+0.32}$~\mearth\ orbiting every $2.3$ days around its host star. Its composition is consistent with an Earth-like planet, and it adds up to the rocky population of sub-Neptunes lying below the density gap identified around M dwarfs. With a $\sim 13$\% precision in mass, a $\sim 7$\% radius precision, and a warm equilibrium temperature of \teq $=543$~K, TOI-771 b is a particularly interesting target for atmospheric characterisation with {\it JWST}, and it is indeed one of the targets under consideration for the Rocky World DDT programme. 
   Additionally, we discover the presence of a second, non-transiting planet in the system, TOI-771 c, with a period of $7.61$ days and a minimum mass of \mplanet~$\sin{i} = 2.87_{-0.38}^{+0.41}$~\mearth. Even though the inclination is not directly constrained, the planet likely belongs to the temperate sub-Neptune population, with an equilibrium temperature of $\sim 365$~K.}
  {}

\titlerunning{TOI-771: a super-Earth and a non-transiting sub-Neptune}
\authorrunning{Lacedelli et al.}
   \maketitle
%
%-------------------------------------------------------------------
\section{Introduction}\label{sec:intro}
%--------------------------------------------------------------------
One of the major discoveries of dedicated exoplanet transit missions like \textit{Kepler} \citep{Borucki2010} and \textit{TESS} \citep{ricker2014} has been the identification of the so-called sub-Neptune population. 
This class of small ($1$~\rearth~$<$~\rplanet~$< 4$~\rearth), close-in ($P < 100$~d) planets is absent in our Solar System, but is likely the most abundant in the Milky Way \citep{Batalha2013, Petigura13PNAS, bean21, biazzo2022}. 
Despite the increasing number of discovered sub-Neptunes, the nature of this population is currently highly debated, both in terms of formation, evolution, and composition. 
In fact, up to now the main observables to constrain their composition have been only mass and radius, but the vast majority of sub-Neptunes lies in a highly degenerate region of the mass-radius diagram, where different theories can explain their bulk densities \citep{RogersSeager10}.
The theories explored till now have mainly followed two strands, namely (1) atmospheric mass-loss processes, based on the atmospheric evolution of rocky cores surrounded by H-He envelopes (e.g. \citealt{OwenWu13, ginzburg2018, Rogers23}), and (2) the water-world hypothesis, with a formation beyond the ice line implying a high mass fraction content of volatiles and subsequent inner migration (e.g. \citealt{Izidoro22, Bitsch19, Venturini20, DornLichtenberg21, burn2021}). 

Both theories can explain the observed properties of the sub-Neptune population \citep{Lissauer11, Fabrycky14, Weiss14, Mulders18,Otegi20}, with one of the most noticeable ones being the dearth of planets in the radius distribution around $1.5$~\rearth, a feature known as the `radius gap' \citep{Fulton2017, FultonPetigura18, cloutier2020, Petigura22}.
Planets below the radius gap are usually termed super-Earths, with bulk compositions consistent with the Earth's.
Planets above the radius gap have lower bulk densities, and they are usually referred to as `mini-Neptunes', even though they can be also called `gas dwarfs' (e.g. \citealt{Buchhave2014, phillips21, Rigby_2024}) or `water worlds' (e.g. \citealt{Leger04, Luque22, Diamond-Lowe22,osborne2023, Cadieux2024a}) depending on the adopted composition and evolution theory.

The launch of \textit{JWST} provided a unique opportunity to start breaking the degeneracy on the composition of sub-Neptunes, providing further observables and using atmospheric studies to investigate their chemistry and interior structures. 
\textit{JWST} recently provided initial, promising results, with the detection of the first molecular features in the atmospheres of a collection of sub-Neptunes (e.g. \citealt{Madhusudhan_2023, bennecke2024, Piaulet2024}). 
In this context, planets orbiting M dwarfs are particularly interesting targets, as M dwarfs are the most numerous stars in our Galaxy, and their low mass and radius make the detection of the planetary atmospheres
less challenging than for Sun-like stars.

Still, to perform accurate atmospheric retrieval, precise mass measurements ($<20$\%) are needed \citep{batalha19}, and a large, homogeneous sample of well-characterised sub-Neptunes is essential for demographic studies. The \texttt{THIRSTEE} project represents a significant step in this direction. This international observing programme aims to shed light on the sub-Neptune population by providing an enlarged sample of precisely characterised planets across stellar types, and combining precise and accurate densities with atmospheric characterization and a novel statistical approach \citep{lacedelli2024}. 
We report here the characterization of the planetary system orbiting one of the \texttt{THIRSTEE} M-dwarf targets, TOI-771.
We analysed \tess\ (Sect.~\ref{sec:tess}) and ground-based photometry together with ESPRESSO radial velocities (RVs) collected under the \thirstee\ programme (Sect.~\ref{sec:ground_based}) to derive stellar parameters (Sect.~\ref{sec:star}) and precise mass, radius, and orbital configuration of the planetary system orbiting TOI-771 (Sect.~\ref{sec:global_analysis}), identifying a second, non-transiting candidate.
Finally, we discuss our results and the characterization of the TOI-771 system in the context of the \thirstee\ project and of the atmospheric characterisation picture (Sect.~\ref{sec:discussion}). 

%--------------------------------------------------------------------
\section{TESS photometry}\label{sec:tess}

\begin{figure}
\centering
\includegraphics[width=\linewidth]{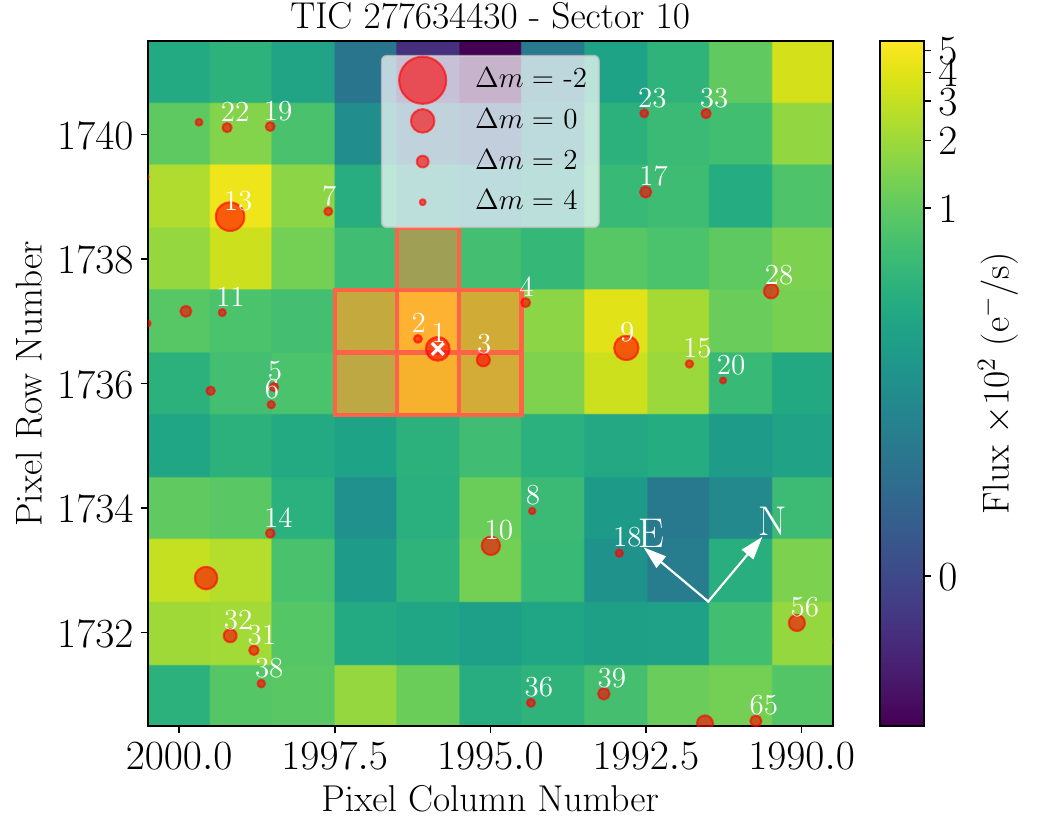}\caption{Sector 10 TPF image of TOI-771 computed with \texttt{tpfplotter}. The red squares mark the \tess\ aperture used for the photometric extraction. Red circles show {\it Gaia} DR3 \citep{GaiaColl2023} stars with magnitude contrast up to $\Delta m = 4$. Star 1 corresponds to TOI-771. The SPOC pipeline accounts for dilution due to stars 2 ($G =16.56 $~mag) and 3 ($G = 15.07 $~mag) using the crowding metric and subtracting the corresponding median flux level percentage \citep{guerrero2021}. The TPFs images of the remaining sectors are reported in Appendix~\ref{appendix:tpf}.}\label{fig:tess_tpf}
\end{figure}

{\it TESS} observed \starname\ for a total of seven sectors, summarised in Table~\ref{tab:TESS_obs}.
The two-minute cadence light curves were reduced using the Science Processing Operations Center (SPOC) 
pipeline \citep{jenkins2016,jenkins2020}. 
The SPOC pipeline identified a $2.33$-day period candidate (TOI-771.01), which was vetted and statistically validated as TOI-771 b by \citep{mistry2024} using the validation tool \texttt{TRICERATOPS} \citep{giacalone2020, Giacalone2021}, together with ground-based photometric and high-resolution imaging follow-up observations from the \tess\ Follow-up Observing Program (TFOP\footnote{\url{https://tess.mit.edu/followup}}; \citealt{collins2019}). 

For our analysis we downloaded the two-minute cadence Pre-search Data Conditioning Simple Aperture Photometry
\citep[PDCSAP, ][]{smith2012, Stumpe2012, Stumpe2014} light curves from the
Mikulski Archive for Space Telescopes (MAST)\footnote{
\url{https://mast.stsci.edu/portal/Mashup/Clients/Mast/Portal.html}}.
The target field of view and the photometric apertures selected for each sector and used by the SPOC pipeline for the light curve extraction are shown in the target pixel file (TPF) images computed with \texttt{tpfplotter}\footnote{\url{https://github.com/jlillo/tpfplotter}.} \citep{Aller2020} in Figs.~\ref{fig:tess_tpf} and \ref{fig:tess_tpf_all}. 

For our photometric analysis, we first removed from the PDCSAP light curves all points flagged as `bad quality' (\texttt{DQUALITY>0})\footnote{ \url{https://archive.stsci.edu/missions/tess/doc/EXP-TESS-ARC-ICD-TM-0014.pdf}}, and we performed a $3 \sigma$ clipping to remove outliers, after masking the in-transit points according to the \cite{mistry2024} planetary parameters.
The resulting light curves are presented in 
Figs.~\ref{fig:tess_lc} and ~\ref{fig:tess_lc_detrended}.

\begin{table}
\caption{Summary of \textit{TESS} observations for TOI-771.}
\small
\label{tab:TESS_obs}
\centering                                      % used for centering table
\begin{tabular}{l l}          % centered columns 
\hline\hline                        % inserts double horizontal lines
Sectors & Dates \\
\hline                                 % inserts single horizontal
10, 11, 12 & March 26-June 19, 2019 \\
37, 38 & April 2-May 26, 2021 \\
64, 65 & April 6-June 2, 2023\\
\hline
\end{tabular}
\end{table}

%--------------------------------------------------------------------
\section{Ground-based follow-up observations}\label{sec:ground_based}

TOI-771 b has been statistically validated by \cite{mistry2024} using ground-based follow-up observations, including high-resolution imaging data and photometric transit observations scheduled using the {\tt TESS Transit Finder} tool \citep{jensen2013} as part of the TFOP programme. 
All the TFOP light curves initially employed in \cite{mistry2024} (their Table 3) as input for the statistical validation of the planet  are available on the {\tt EXOFOP-TESS} webpage\footnote{\url{https://exofop.ipac.caltech.edu/tess/target.php?id=277634430}}. 
Here, we included all those light curves for the first time in the global modelling (Sect.~\ref{sec:joint_fit}).
In addition, we collected four additional transits as part of the SPECULOOS consortium.
We detail below all the ground-based photometric and RV observations that we employed in this work to characterise TOI-771 b. 

\subsection{M-Earth-South photometry}\label{sec:mearth}

We observed one transit window of TOI-771 b as part of TFOP with the eight 0.4 m M-Earth-South telescopes \citep{irwin2007} located at Cerro Tololo Inter-American Observatory, Chile.
Observations were taken on February 2, 2020 using the RG715 filter. 
Data analysis and extraction were performed through the custom pipelines outlined in \cite{irwin2007}.
We used a circular photometric aperture of $5\farcs 0$, which excluded the nearest neighbour in the {\it Gaia} DR3 catalogue ({\it Gaia} DR3 5229384714548082944).

\subsection{TRAPPIST-South photometry}\label{sec:trappist}

We observed two transit windows of TOI-771 b as part of TFOP with the TRAnsiting Planets and PlanetesImals Small Telescope (TRAPPIST) South 0.6 m telescope \citep{jehin2011, gillon2011} located at ESO La Silla Observatory, Chile. 
The first transit was observed on January 29, 2022 in the I+z band, while the second one on February 2, 2022 in the Sloan $z'$ band. 
As part of the TRAPPIST consortium, we collected an additional transit on March 19, 2022 in the I+z band.
Data calibration and photometric extraction were performed using either \texttt{AstroImageJ} or the dedicated \texttt{prose} pipeline outlined in \cite{garcia2022}\footnote{\url{https://github.com/lgrcia/prose}.}.
We used circular photometric apertures of $4\farcs 5$, $5\farcs 0$, and $4\farcs 6$ respectively, to exclude the nearest neighbour star {\it Gaia} DR3 5229384714548082944. \\

\subsection{SPECULOOS-South}\label{sec:lcogt}

We observed two transit windows of TOI-771-b as part of TFOP with the SPECULOOS Southern Observatory (SSO, \citealt{jehin2018}) located at the ESO Paranal Observatory, Chile. 
Both transits were observed in the Sloan $g'$ band, one with the 1-m SSO-Io telescope on February 5, 2022, and one with the 1-m SSO-Europa telescope, on April 2, 2022. 
As part of the SPECULOOS consortium, we collected three additional transits, one on March 19, 2022 with the SSO-Europa telescope in the Sloan-$g'$ band, and two during the same night on April 20, 2024, one with the SSO-Io telescope and one with the SSO-Callisto telescope, both in the Sloan-$i'$ band.
Data calibration and extraction was performed with a specific pipeline, reported in \cite{sebastian2020}.
We used circular photometric apertures of $3\farcs 3$ for the SSO-Io and $3\farcs 6$ for SSO-Europa TFOP observations, to exclude the nearest neighbour star {\it Gaia} DR3 5229384714548082944.
For the additional transits, circular photometric apertures of $2\farcs 2$, $5\farcs 1$, and $3\farcs 0$ were used for the Europa, Io, and Callisto telescopes, respectively. 

\subsection{Long-term photometric monitoring}\label{sec:asas}
\begin{figure*}
\centering
  \includegraphics[width=\linewidth]{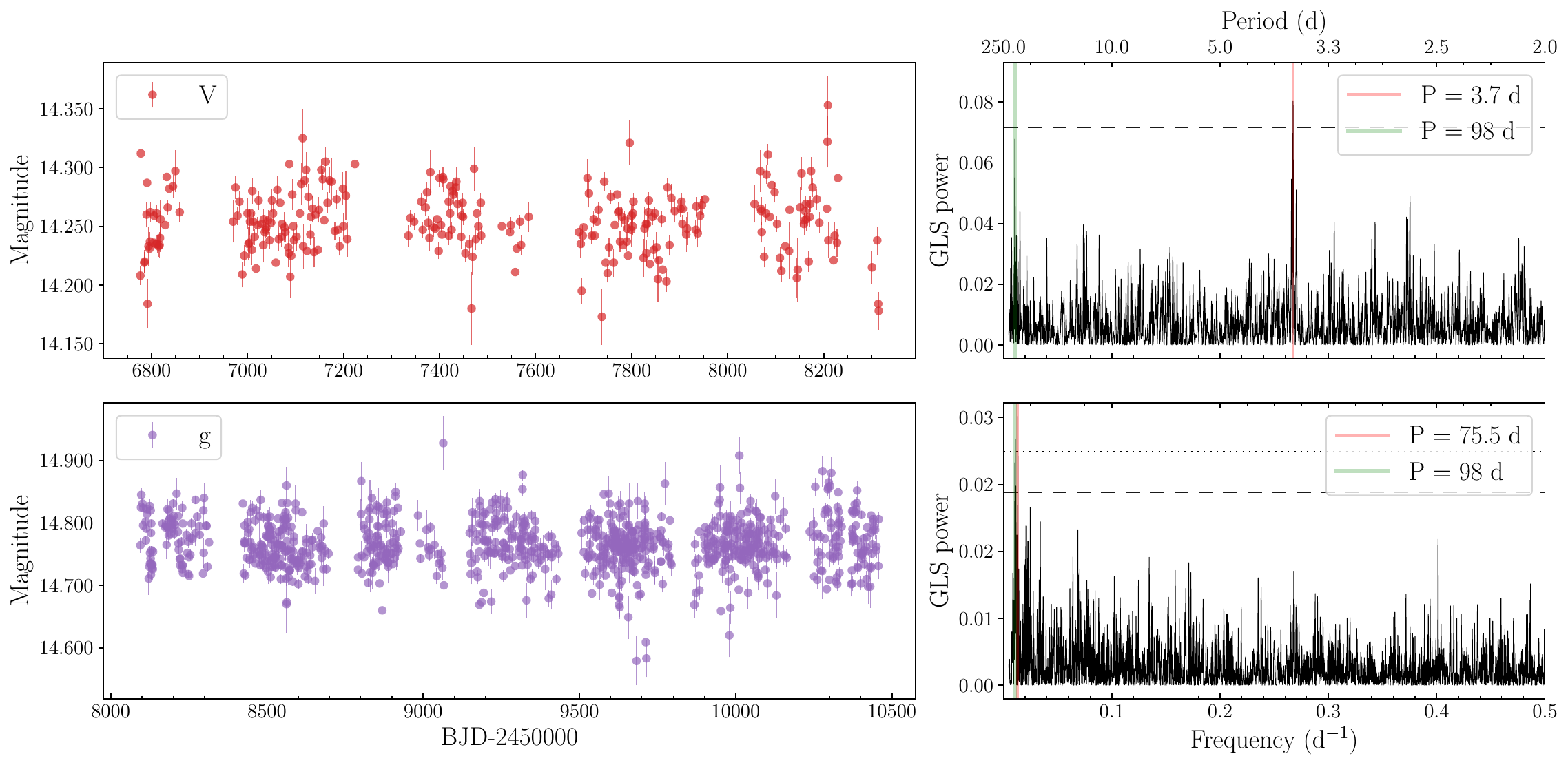}
  \caption{ASAS-SN {\it V}- and {\it g} photometry of TOI-771. The GLS periodogram of each light curve is shown in the right panel, with the $10$\% and $1$\% false alarm probability (FAP) levels marked with an horizontal dashed and dotted line, respectively. The vertical red line highlights the most significant peak, while the vertical green line shows the periodicity at $ \sim 98$~d.
  }\label{fig:asas_sn}
\end{figure*}

We recovered the All-Sky Automated Survey for Supernovae (ASAS-SN; \citealt{Shappee2014}, \citealt{Kochanek2017}) photometric {\it V} and {\it g} long-term light curves of TOI-771 from the ASAS-SN portal\footnote{\url{http://asas-sn.ifa.hawaii.edu/skypatrol/}.} \citep{hart2023}. 
The star was observed in the {\it V} filter between 2014 and 2018, for a total time span of $\sim 1538$ days, with a median cadence of $\sim 3$ days. 
The median magnitude is $V = 14.25$~mag, the median photometric uncertainty $\sigma_{V}$ is $0.009$ mag, and the root mean square (RMS) is $0.027 $~mag. 
In the {\it g} filter, observations span from 2017 to 2024, for a total of $\sim 2363$ days, with a median cadence of $\sim 1$~d. The median magnitude, uncertainty, and RSM are $g = 14.77$~mag, $\sigma_{g} = 0.010$ mag, and RMS~$= 0.039$~mag. The time series are plotted in Fig.~\ref{fig:asas_sn}.  
The photometric precision and cadence of the ASAS-SN light curves are not suited to detect the transiting planet, but they were employed to study the stellar activity and investigate the stellar rotational period (see Sect.~\ref{sec:star_activity}). 
For this analysis, we filtered the light curves selecting only the good-quality flagged observations (\texttt{QUALITY=G}), and rejecting the outliers by applying a $5 \sigma$ clipping, as in \citep{hart2023}.

%--------------------------------------------------------------------
\subsection{ESPRESSO spectroscopic observations}\label{sec:spectra}

We collected $31$ high-resolution spectra of TOI-771 with the Echelle SPectrograph for Rocky Exoplanets and Stable Spectroscopic Observations (ESPRESSO, \citealt{pepe2021}) installed at the ESO Very Large Telescope telescope array at Paranal Observatory, Chile. 
The star was selected as part of \thirstee\ RV follow-up \citep{lacedelli2024}, and it was observed within the 112.25F2.001 programme (PI: E. Palle).
The observations span between December 6, 2023, and March 9, 2024, covering a total baseline of $\sim 94$ days.
We employed the ESPRESSO’s single Unit Telescope (1UT) set-up, using the high-resolution (HR) mode ($1$~arcsec fibre, $R \sim 140 000$) with the $2 \times 1$ detector binning (HR21), covering the spectral range $\sim 380 - 780$~nm.
For background subtraction, we selected the observing configuration with fibre B placed on the sky.
Spectra were gathered with an exposure time of $1800$ seconds, resulting in a median signal-to-noise ratio (\snr) of $17$ at $550$~nm. 
Data were reduced using ESPRESSO's data reduction pipeline (DRS), version 3.0.0. 
Following the methodology of \cite{lacedelli2024}, seeking for an homogeneous analysis of the whole \thirstee\ sample, we extracted the RV values using the template-matching  \texttt{SERVAL}\footnote{\url{https://github.com/mzechmeister/serval}.} algorithm \citep{zechmeister2018}. 
The resulting RV dataset, listed in Table~\ref{table:RV_table_ESPRESSO}, has an RMS of $3.99$~\ms\ and a median internal uncertainty of $0.65$~\ms. 

%--------------------------------------------------------------------
\section{The star}\label{sec:star}

\begin{table}[h!]
\small
\caption{Stellar properties of TOI-771.}
\label{table:star_params}
\begin{threeparttable}[t]
\centering
\begin{tabular}{lll}
\hline\hline
\multicolumn{3}{c}{\starname}\\
\hline
TIC & \multicolumn{2}{l}{277634430} \\
% TYC & \multicolumn{2}{l}{243-1528-1}\\
{\it Gaia} DR3 & \multicolumn{2}{l}{5229384714547758720}\\
2MASS & \multicolumn{2}{l}{J10562716-7259054} \\
% LP & \multicolumn{2}{l}{994-91}\\[1ex]
\hline
Parameter & Value & Source \\
\hline
RA  (J2000; hh:mm:ss.ss) &  10:56:27.34 & A\\
Dec (J2000; dd:mm:ss.ss) & $-$72:59:06.66 & A\\
$\mu_{\alpha}$ (mas yr$^{-1}$) & $39.300 \pm 0.017 $ & A\\
$\mu_{\delta}$ (mas yr$^{-1}$) & $-76.417 \pm 0.015$ & A\\
Parallax (mas) & $39.4467 \pm 0.0136$& A\\
Distance (pc) & $25.325 \pm 0.008$& B \\
$\gamma$ (\kms) & $-5.90 \pm 1.02$ & A \\
U (\kms) & $ 4.56 \pm 0.42 $& F$^a$  \\ 
V (\kms) & $ 9.86 \pm 0.91 $& F$^a$  \\
W (\kms) & $ -4.82 \pm 0.21$& F$^a$  \\
\hline
{\it TESS} (mag) & $12.087 \pm 0.007$& C\\
{\it G} (mag) & $13.343 \pm 0.003$ & A\\
{\it G$_{\rm BP}$} (mag) & $14.877 \pm 0.004$ & A\\
{\it G$_{\rm RP}$} (mag) & $12.141\pm 0.004$ & A\\
{\it B } (mag) & $16.768 \pm 0.162$ & C\\
{\it V }(mag) & $14.888 \pm	0.080$ & C\\
{\it J }(mag) & $10.507  \pm 0.023$ & D\\
{\it H }(mag) & $9.955  \pm0.022$ & D\\
{\it K }(mag) & $9.664  \pm	0.021$ & D\\
{\it W1} (mag) & $9.479 \pm	0.023$ & E\\
{\it W2} (mag) & $9.303  \pm 0.020$ & E\\
{\it W3} (mag) & $9.173 \pm	0.024$ & E\\
{\it W4} (mag) & $8.885 \pm	0.216$ & E\\[1ex]
\hline
\teff\ (K) & $ 3370 \pm 100 $ & F \\ 
\logg\  (cgs)  & $4.80  \pm 0.07$ & F, {\tiny Spectroscopic} \\
\logg\  (cgs)  & $ 5.05 \pm 0.14$ & F, {\tiny Bolometric} \\ 
$[$Fe/H$]$ (dex) & $ -0.13 \pm 0.07$ & F \\
\logRHK & $  -6.36 \pm 0.06  $ & F\\
$P_{\rm rot}$ (d) & $ 98_{-5}^{+10}$ & F\\
\rstar\ (\rsun) & $ 0.232^{+0.017}_{-0.023}$ & F\\
\mstar\ (\msun) & $  0.220^{+0.030}_{-0.035}$ & F \\
$L_{\star}$ ($L_{\odot}$) & $ 0.00626^{+0.00015}_{-0.00053} $ & F\\ 
Spectral type & M3 & F\\
\hline\hline
\end{tabular}
\tablebib{
\small
A) {\it Gaia} DR3 \citep{GaiaColl2023}. B) \citet{bailer_jones2021}.
C) {\it TESS} Input Catalogue Version 8 (TICv8, \citealt{Stassun2018}).
D) Two Micron All Sky Survey (2MASS, \citealt{Cutri2003}).
E) {\it Wide-field Infrared Survey Explorer} \citep[{\it AllWISE};][]{cutri_allwise}. F) This work. \\
\tablefoottext{a}{Space velocity components in the Galactic, heliocentric, right-handed system.}}
\end{threeparttable}
\end{table}

\subsection{Stellar parameters}\label{sec:star_properties}

To obtain homogeneous parameters for the \texttt{THIRSTEE} sample, we used the same methodology as in \citet{lacedelli2024} to derive the stellar parameters of TOI-771. 
We applied the {\sc SteParSyn}\footnote{\url{https://github.com/hmtabernero/SteParSyn/}} \citep{tab22} code to the ESPRESSO spectra, combining the BT-Settl stellar atmospheric models \citep{all12}, and atomic and molecular data from the Vienna atomic line database \citep[VALD3, see][]{rya15} with the Turbospectrum-generated \citep{ple12} grid of synthetic spectra. 
We employed an optimised set of \ion{Fe}{i} and \ion{Ti}{i} lines, as well as various TiO molecular bands, for M-type stars, following \citep{mar21}. 
We obtained the following parameters: $T_{\rm eff}$~$=$~3370~$\pm$~18~K, $\log{g}$~$=$~4.80~$\pm$~0.07 dex, and [Fe/H]~$=$~$-0.13$~$\pm$~0.07~dex. 
The first set of error bars refers only to the internal errors, and to estimate a more realistic uncertainty on \teff (see also \citealt{tayar2022}), we added a systematic component to the errors, following \cite{mar21}, for a final temperature value and uncertainties of $T_{\rm eff}$~$=$~3370~$\pm$~100~K.
All our derived parameters, listed in Table~\ref{table:star_params}, are consistent within $1 \sigma$ with \cite{mistry2024}.

\begin{figure}
\centering
  \includegraphics[width=\linewidth]{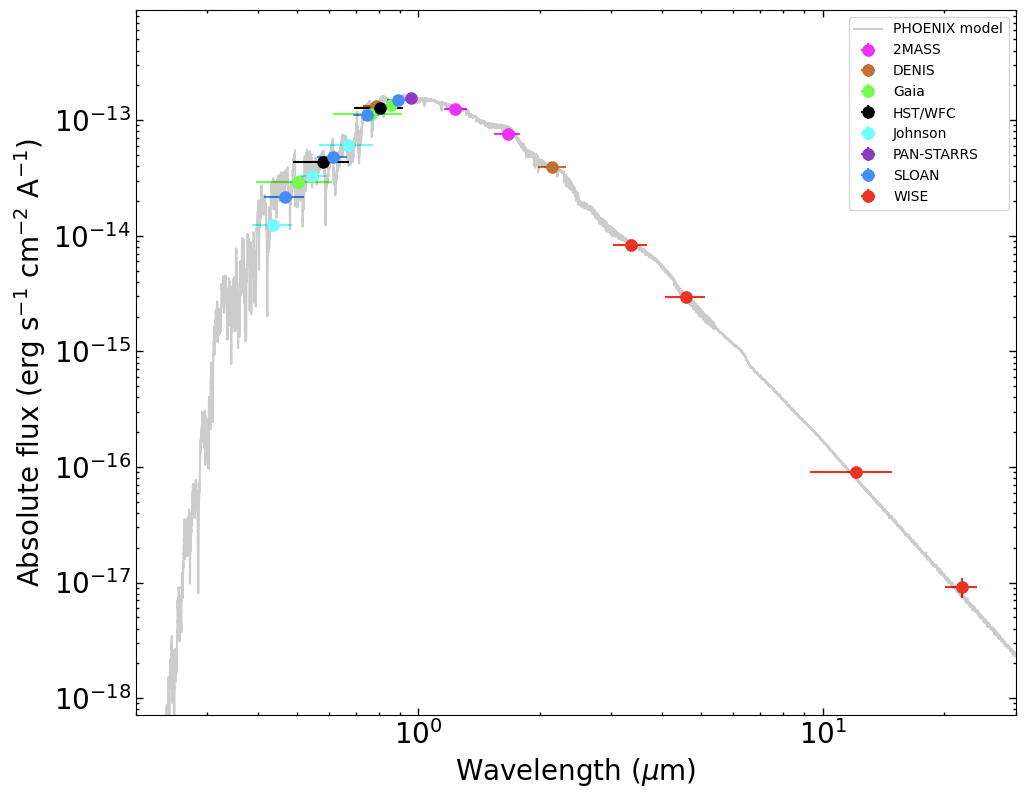}
  \caption{SED of TOI-771. The grey line shows the BT-Settl model (which uses the PHOENIX atmospheric code) assuming \teff~$=3300$~K, \logg~$= 5.0$~dex, and solar metallicity. 
  }
    \label{fig:sed}
\end{figure}

To derive the stellar luminosity, mass, and radius, we first computed the photometric spectral energy distribution (SED) of TOI-771 (Fig.~\ref{fig:sed}) using the publicly available broad-band photometry from
the American Association of Variable Star Observers Photometric All-Sky Survey \citep[APASS,][]{henden_2014}, 
the Sloan Digital Sky Survey \citep[SDSS instead][]{york_2000}, the {\it Gaia} DR3 archive \citep{gaia_2016,GaiaColl2023},  2MASS \citep{skrutskie_2006}, and the Wide-field Infrared Survey Explorer archive \citep[{\sl WISE},][]{wright_2010}.
We used the zero points listed in the Virtual Observatory SED Analyzer tool \citep[VOSA,][]{bayo2008} to convert the observed magnitudes into fluxes, and the {\it Gaia} DR3 trigonometric parallax to transform from observed to absolute fluxes.  
We then integrated the observed SED to obtain the stellar bolometric luminosity \hbox{$L_{\star} = 0.00626^{+0.00015}_{-0.00053}$~L$_\odot$}, which we employed together with our derived \teff\ to compute the stellar radius ($R_{\star} = 0.232 ^{+0.017}_{-0.022}$~\rsun) using the Stefan-Boltzmann law.
We then used the mass-radius relation of \cite{schweitzer2019} to derive the stellar mass $M_{\star} = 0.221^{+0.030}_{-0.035}$~\msun.
As a consistency check, we compared the surface gravity obtained using our mass and radius values (\logg$= 5.05 \pm 0.14$~dex), with our derived spectroscopic value (log\,$g = 4.80 \pm 0.07$~dex), confirming that they are consistent within 1.5\,$\sigma$. 

Finally, we performed a kinematic analysis of TOI-771 to assess its position in the Galactic framework. Starting from {\it Gaia} DR3 positions, proper motions, parallax, and systemic RV (see Table~\ref{table:star_params}), we computed the galactic heliocentric space velocities in the directions of the Galactic centre, Galactic rotation, and north Galactic pole, obtaining $U = 44.05 \pm 0.31$ \kms, $V = -44.44 \pm 0.29 $ \kms, and $W = -0.18 \pm 0.29 $ \kms, respectively.
Based on these values, and adopting the probabilistic framework of \cite{bensby2014}, we infer that TOI-771 kinematically belongs to the thin-disk population. 

\subsection{Activity indices and rotational period}\label{sec:star_activity}

To identify the stellar rotational period, we first analysed the ASAS-SN photometry.
The generalised Lomb-Scargle (GLS, \citealt{Zechmeister2009}) periodogram of the ASAS-SN $V$ and $g$ light curves show the most significant periodicity at $3.7$~d and $75.5$~d, respectively (see Fig.~\ref{fig:asas_sn}). 
However, the $3.7$~d periodicity is not identified in the {\it TESS} periodogram, neither in the PDCSAP nor in the SAP light curves, and it is likely to be related to the light curve cadence ($\sim 3$~d). 
Additionally, both ASAS-SN time series show a second periodicity around $\sim 98$~d, especially significant in the {\it g} filter. We searched for this long-period signal in the TESS SAP light curve, but we could not identify it, probably due to the clustering of the sectors, favouring shorter periods (see Appendix~\ref{appendix:light_curves} and Fig.~\ref{fig:tess_sap}).

This long-term periodicity is also identified in the spectroscopic time series. Indeed, we analysed various spectral activity indicators of the ESPRESSO dataset, including the  \texttt{SERVAL} absorption line indicators (Na \textsc{i} doublet ($\lambda \lambda 589.0$~nm, $589.6$~nm), \halpha\ ($\lambda 656.2$~nm), and $S$-index (Ca \textsc{ii} H\&K lines, $\lambda\lambda 396.8470$~nm, $393.3664$~nm), as derived from the \texttt{ACTIN2}\footnote{\url{https://github.com/gomesdasilva/ACTIN2}} code \citep{Gomes2018, Gomes2021}), the RV chromatic index (CRX), and the differential line width (dLW; see \citealt{zechmeister2018} and \citealt{jeffers2022} for more details). 
Moreover, we included in the analysis the ESPRESSO DRS \citep{Baranne1996, Pepe2002} activity indexes, namely the bisector span (BIS) \citep{queloz2001}, the full-width-half-maximum (FWHM), and the contrast (Contr) of the cross-correlation function (CCF), which was computed using the {\tt M3} ESPRESSO DRS mask.

\begin{figure}[h!]
\centering
  \includegraphics[width=\linewidth]{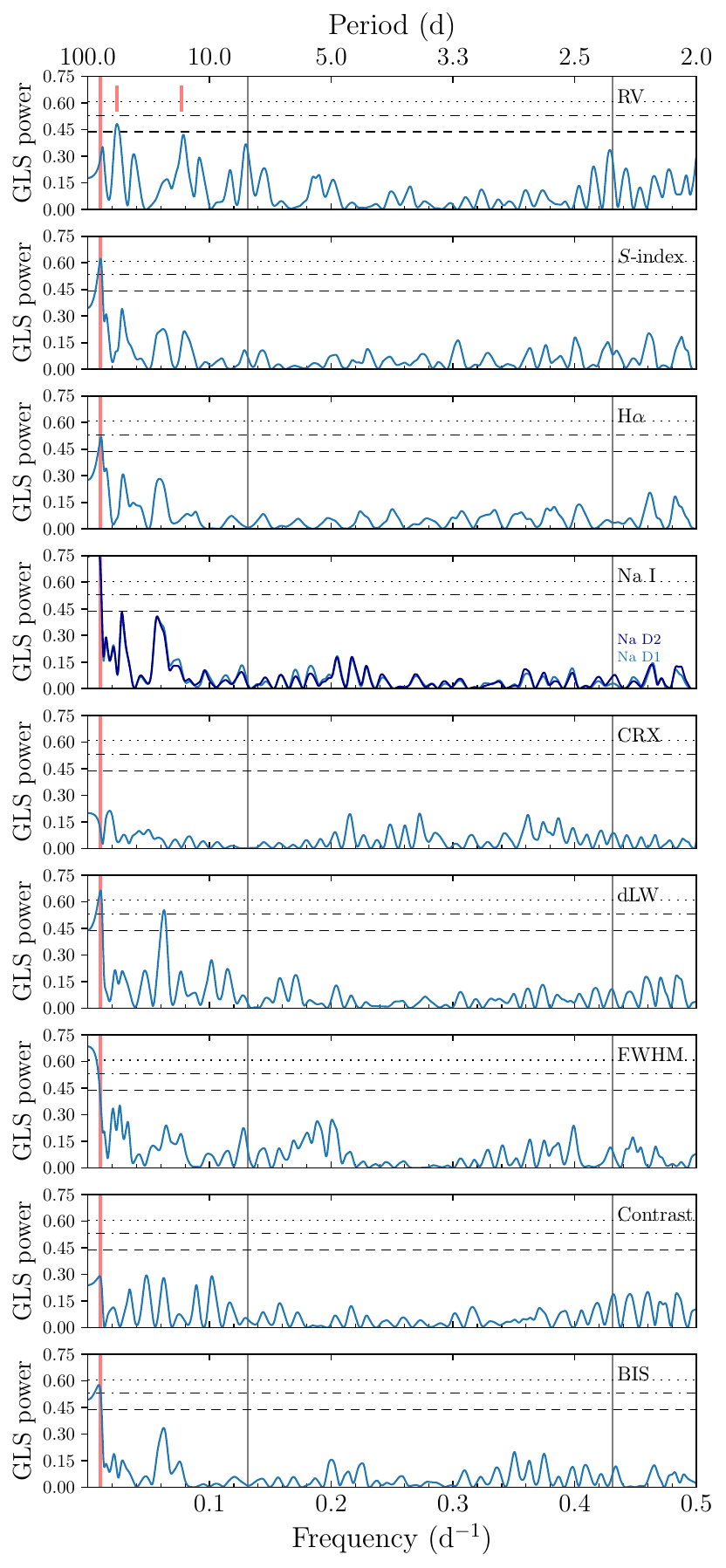}
  \caption{GLS periodogram of the TOI-771 ESPRESSO RVs and activity indicators. The vertical red line marks the possible stellar rotational period at $\sim 98$ days. The vertical red ticks in the RV periodogram mark the peaks at $\sim 42$~d and $\sim 13$~d, which are related to each other, and to the $98$-d signal (see Fig.~\ref{fig:iterative_periodogram}). The vertical grey lines show the periods of TOI-771 b ($2.33$~d), and of the second planet candidate TOI-771 c ($\sim 7.6$~d). The $10\%$, $1$\%, and 0.1\% FAP levels are drawn with horizontal dashed, dash-dotted, and dotted lines, respectively.
  }
    \label{fig:GLS_indicators_ESPRESSO}
\end{figure}

\begin{figure}
\centering
  \includegraphics[width=\linewidth]{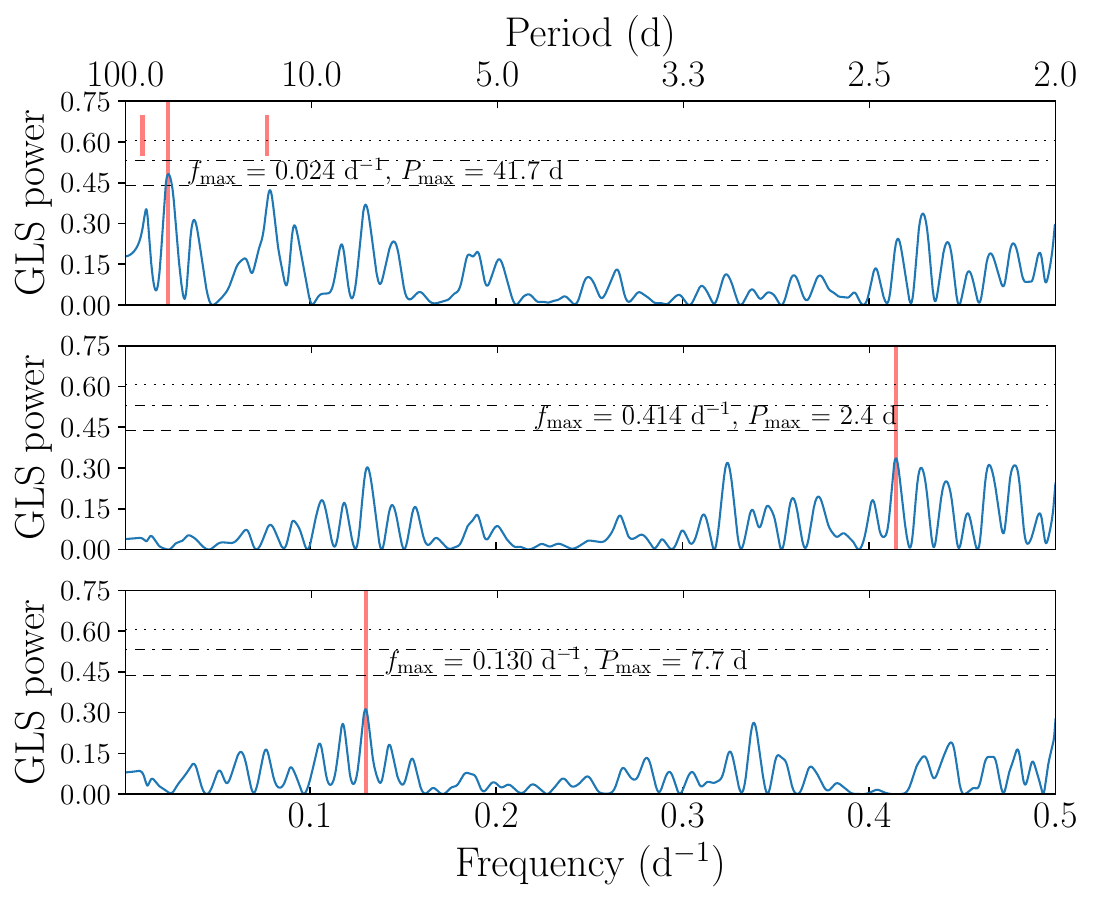}
  \caption{Iterative GLS periodogram of the TOI-771 ESPRESSO RVs. The sinusoidal model corresponding to the best frequency is subtracted at each iteration, starting from top to bottom. The vertical red line marks most significant peak in each iteration, and the corresponding frequency and period are reported. 
  The vertical red ticks in the first panel mark the peaks at $\sim 98$~d and $\sim 13$~d, which are related to each other and to the $42$-d signal. All three peaks disappear after the first iteration.
  The $10\%$, $1$\%, and 0.1\% FAP levels are drawn with horizontal dashed, dash-dotted, and dotted lines, respectively.
  }
    \label{fig:iterative_periodogram}
\end{figure}

We show the GLS periodograms of the ESPRESSO RVs and activity indexes in Fig.~\ref{fig:GLS_indicators_ESPRESSO}.  
The most significant peak in the RVs periodogram is identified at $\sim 42$~d, and we associate it with an alias of the long-term signal at $\sim 98$~d. 
In fact, when removing the $42$~d signal from the periodogram, the $\sim 98$~d peak also vanishes (see Fig.~\ref{fig:iterative_periodogram}). 
The $98$-d signal is also present with high significance in various activity indicators, especially in the $S$-index, \halpha, dLW, and BIS (Fig.~\ref{fig:GLS_indicators_ESPRESSO}).

To search for additional periodicities in the RV time series, we performed an iterative frequency search on the GLS periodogram, subtracting at each iteration the sinusoidal model associated with the most significant frequency. 
As Fig.~\ref{fig:iterative_periodogram} shows, our search identified the frequencies $f_{\rm 1} = 0.024$~d$^{-1}$
($P_{\rm 1} = 41.8$~d), $f_{\rm 2} = 0.415$~d$^{-1}$
($P_{\rm 2} = 2.4$~d, corresponding to TOI-771 b), and $f_{\rm 3} = 0.130$~d$^{-1}$
($P_{\rm 3} = 7.7$~d), even though with low significance. 
The first signal at $42$~d is related not only to the $98$-d peak, but also to the one at $\sim 13$~d, since they both disappear in the periodogram when subtracting the $42$-d signal (Fig.~\ref{fig:iterative_periodogram}).

\begin{figure}
\centering
  \includegraphics[width=\linewidth]{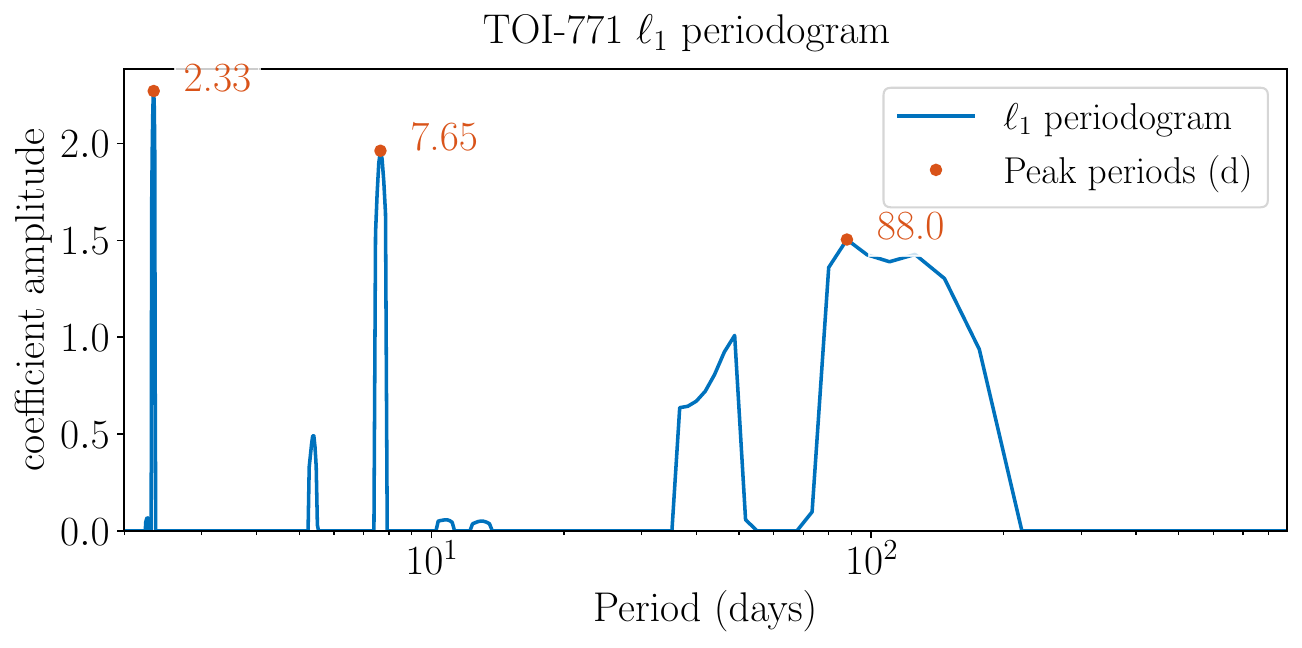}
  \caption{$\ell_1$ periodogram of TOI-771 ESPRESSO RVs. The frequency grid is computed from 0 to 1 cycle per day. The time-span of the observations is $\sim 94$ days. Significant peaks (with $\Delta\ln\mathcal{Z} > 10$, $\mathcal{Z}$ being the Bayesian evidence) are highlighted with orange dots.
  }
    \label{fig:l1_periodogram}
\end{figure}

The presence of these three peaks, with the long-term periodicity recovered as a broad peak around $100$~d, TOI-771 b' signal at $\sim 2.3$~d, and an additional signal at $\sim 7.7$~d, is further confirmed from our analysis using the $\ell_1$ periodogram\footnote{\url{https://github.com/nathanchara/l1periodogram}.} \citep{hara2017}.
Due to the intrinsic algorithm construction, fewer peaks related to aliases are identified in the $\ell_1$ periodogram (Fig.~\ref{fig:l1_periodogram}).
Given the presence of a similar periodicity (around $90-100$~d) in the ASAS-SN photometry, as well as in various spectroscopic activity indicators, we identify this long-term signal as the rotational period of the star, and we fitted the RVs time series together with the S-index (see Sec.~\ref{sec:joint_fit}) to recover the stellar rotational period, obtaining \prot $ = 98_{-5}^{+10}$~d. 
Such a slow rotation is consistent with the low activity level of the star (\logRHK~$= -6.361 \pm  0.056$, computed from the median $S$-index, assuming $B-V = 1.88$ and applying \citep{suarez_mascareno2016} relations) in the optical regime \citep{suarez_mascareno2016, Astudillo-Defru_2017}. 
However, considering the $94$-days time span of the current ESPRESSO dataset, RV observations spanning a longer baseline will be necessary to definitively confirm the nature and periodicity of this signal, which is only broadly constrained by the current analysis.

%--------------------------------------------------------------------
\section{Data analysis and results}\label{sec:global_analysis}
  
\subsection{Photometric fit}\label{sec:photometric_fit_only}

We performed an initial fit of the \textit{TESS} light curves only, to recover the transit of TOI-771 b using \pyorbit\footnote{\url{https://github.com/LucaMalavolta/PyORBIT}.} \citep{Malavolta2016, Malavolta2018}, a versatile Python framework that allows for the modelling of light curves, RVs, transit time variations, and stellar activity. 
We employed the PyDE\footnote{\url{https://github.com/hpparvi/PyDE}} + {\tt emcee} \citep{ForemanMackey2013} \pyorbit\ set-up, using the same convergence criteria of \citep{lacedelli2022}. 
We employed $2 n_{\mathrm{dim}}$ walkers, being $n_{\mathrm{dim}}$ the dimensionality of the model, and we ran the chains for $250000$ iterations, discarding the first $30000$ steps as burn-in and adopting a thinning factor of $100$. 
We used {\tt batman} \citep{Kreidberg2015} as implemented in \pyorbit\ to fit the mid-transit time ($T_0$), the period ($P$), the stellar radius ratio ($R_\mathrm{p}/$\rstar), the impact parameter ($b$), and the stellar density (\rhostar), using our derived parameters (Table~\ref{table:star_params}) as a Gaussian prior.
We adopted a quadratic limb-darkening (LD) law with \cite{Kipping2013} parametrisation ($q_1$, $q_2$), using \texttt{PyLDTk}\footnote{\url{https://github.com/hpparvi/ldtk}.} \citep{Husser2013, Parviainen2015} to estimate the initial values of the coefficients, which we imposed as a Gaussian priors adopting a custom $0.1$ uncertainty. 
Finally, we included a jitter term to account for additional, uncorrelated noise.
For all the fitted parameters, we adopted uniform priors, unless specified otherwise.

We performed an initial fit including a Gaussian process (GP) regression model with a Matérn-3/2 kernel. The resulting model is shown in Fig.~\ref{fig:tess_lc}.
Given the low activity level of the star, and the absence of strong correlated noise, we then tested a second model on pre-detrended \textit{TESS} light curves. We detrended the light curves using \texttt{WOTAN}\footnote{\url{https://github.com/hippke/wotan}.} \citep{hippke2019}, adopting a biweight time-windowed slider with a $1$~d window, and
masking the in-transit points to preserve the transit shape. The resulting light curves and model are shown in Fig.~\ref{fig:tess_lc_detrended}.
A comparison between the planetary parameters derived from both fits provided consistent results. Therefore, to reduce the computational time, for the joint analysis in Sect.~\ref{sec:joint_fit} we adopted the pre-detrended light curves without including a GP model. 

\subsection{Radial velocity fit}\label{sec:RV_fit_only}

According to our frequency analysis, the signal identified in the RV time series at $7.6$~d is unlikely to be related to stellar activity, as none of the activity indicators shows significant peaks at such a period. 
Therefore, we hypothesised the presence of an additional planetary candidate, and we tested its presence by comparing the Bayesian evidence of a one-planet model versus a two-planet model. 
We computed the Bayesian evidence using the dynamical nested-sampling (NS) algorithm \texttt{dynesty} \citep{skilling2004, skilling2006, speagle2020} as implemented in \pyorbit.
In both fits, we assumed for each planet a wide uniform prior on the RV semi-amplitude ($K \in [0.01$-$100$]~\ms), exploring it in logarithmic space, and a half-Gaussian zero-mean prior \citep{vanEylen2019} on the eccentricity $e$, adopting the ($\sqrt{e} \sin{\omega}$, $\sqrt{e} \cos{\omega}$) parametrisation of \cite{Eastman2013}, where $\omega$ is the argument of periastron.
For TOI-771 b, we assumed Gaussian priors on $T_0$ and $P$ from \citep{mistry2024}, and in the two-planet fit we allowed the period of the second candidate to span between $3$ and $40$~days, that is, with the upper limit smaller than the alias of the stellar rotational period.
We also included a systemic RV offset, and a jitter term to account for extra stellar and instrument noise.
We investigated both the cases with and without stellar activity, to test the robustness of the planetary detection.
When including stellar activity, we modelled it using a GP regression with a quasi-periodic kernel \citep{Grunblatt2015}. 
We assumed as GP hyper-parameters the stellar rotation period (\prot), the characteristics decay timescale ($P_\mathrm{dec}$), the coherence scale ({\it w}), and the GP amplitude ($A$).
We ran the NS fits assuming 1000 live points and a sampling efficiency of 0.3. \\
Table~\ref{tab:RV_comparison} reports the obtained Bayesian evidences of the models, as well as the semi-amplitudes, and the period of the second candidate.
Overall, the models including the GP modelling are strongly favoured with respect to the case with no stellar activity, with a difference in the logarithmic Bayes
factor $2 \, \Delta\ln\mathcal{Z} > 7 $ \citep{kass&raftey1995}. 
Moreover, both in the case with and without the GP modelling, the two-planet model is favoured with respect to the one-planet model.
Therefore, also considering our frequency analysis in Sect.~\ref{sec:star_activity}, we decided to adopt the two-planet model with the GP as the final one, and we identify the $7.6$~d periodicity with a second planet in the system, TOI-771 c. 
As Table~\ref{tab:RV_comparison} shows, the semi-amplitude of TOI-771 b is consistent among all fits, but modelling the stellar activity and the additional signal including a second planet improves significantly the precision on $K_{\rm b}$. 

\begin{table}[h!]
\caption{Comparison between the one-planet (1P) and two-planet (2P) models parameters, with and without GP modelling.}
\small
\label{tab:RV_comparison}
\centering                                      % used for centering table
\begin{tabular}{l c c c c}          % centered columns 
\hline\hline                        % inserts double horizontal lines
Model & $\Delta \ln\mathcal{Z}$ & $K_{\mathrm b}$ (\ms) & $K_{\mathrm c}$ (\ms) & $P_{\mathrm c}$ (d)\\
\hline                                 % inserts single horizontal
1P & $-98.9 \pm 0.1$ & $ 2.7_{-1.5}^{+1.0} $ & -  &  - \\
2P &  $-97.4  \pm 0.2  $& $2.99_{-0.75}^{+0.71} $& $ 3.02_{-1.10}^{+0.71}$  &  $7.64_{-0.09}^{+0.07}$\\[1ex]
\hline
1P + GP & $-93.9 \pm  0.1$ & $3.11_{-0.59}^{+0.52}$ & -  &  - \\
2P + GP &    $-89.1 \pm  0.2$& $3.32 \pm 0.32$& $2.55_{-0.56}^{+0.33}$  &  $7.60_{-0.03}^{+0.03}$\\
\hline
\end{tabular}
\end{table}

As a further test, to evaluate the robustness of the obtained semi-amplitude of TOI-771 b, we also tried a less flexible approach, modelling the stellar activity with a sinusoid instead of using GPs. 
We included in the model the two planetary signals plus a third Keplerian with a circular orbit, and we explored the periodicities at $\sim 98$~d and its alias at $\sim 42$~d  in two different runs, using wide, uniform priors around the test period. 
In the first case, we obtained a consistent semi-amplitude of $K_{\mathrm b} = 3.49 \pm 0.31$~\ms, recovering a period of $102_{-7}^{+10}$~d for the sinusoidal signal. 
When fitting the activity period alias (deriving $P_{\rm rot} 
 = 42_{-4}^{+2}$~d), the fit struggled to reach convergence, and we only obtained a partial (yet consistent) detection for planet b ($K_{\mathrm b} = 2.33_{-1.0}^{+0.68}$~\ms). 
Moreover, the difference in the Bayes factor between the two models ($2 \, \Delta\ln\mathcal{Z} = 19.9$) strongly favours the $\sim 98$~d period for the stellar activity modelling.
In all cases, the semi-amplitude of TOI-771 b is consistent within $1 \sigma$ between the models, showing the robustness of the result. 

\subsection{Joint photometric and RV modelling}\label{sec:joint_fit}
\begin{figure}
\centering
  \includegraphics[width=\linewidth]{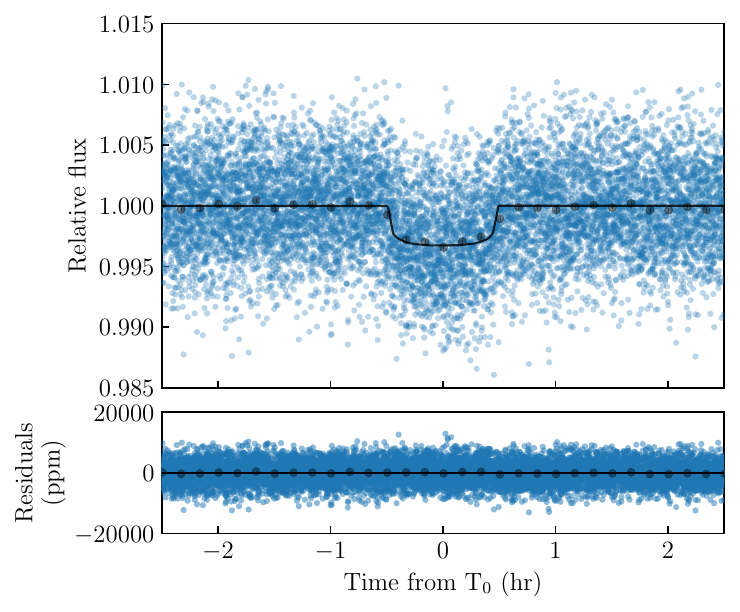}
  \caption{Phase-folded \tess\ photometry of TOI-771 b. The solid black line shows the best-fitting model. Data binned over $10$~min are shown as black dots. Residuals are displayed in the bottom panel. 
  }
    \label{fig:TESS_phase}
\end{figure}

\begin{figure}
\centering
  \includegraphics[width=\linewidth]{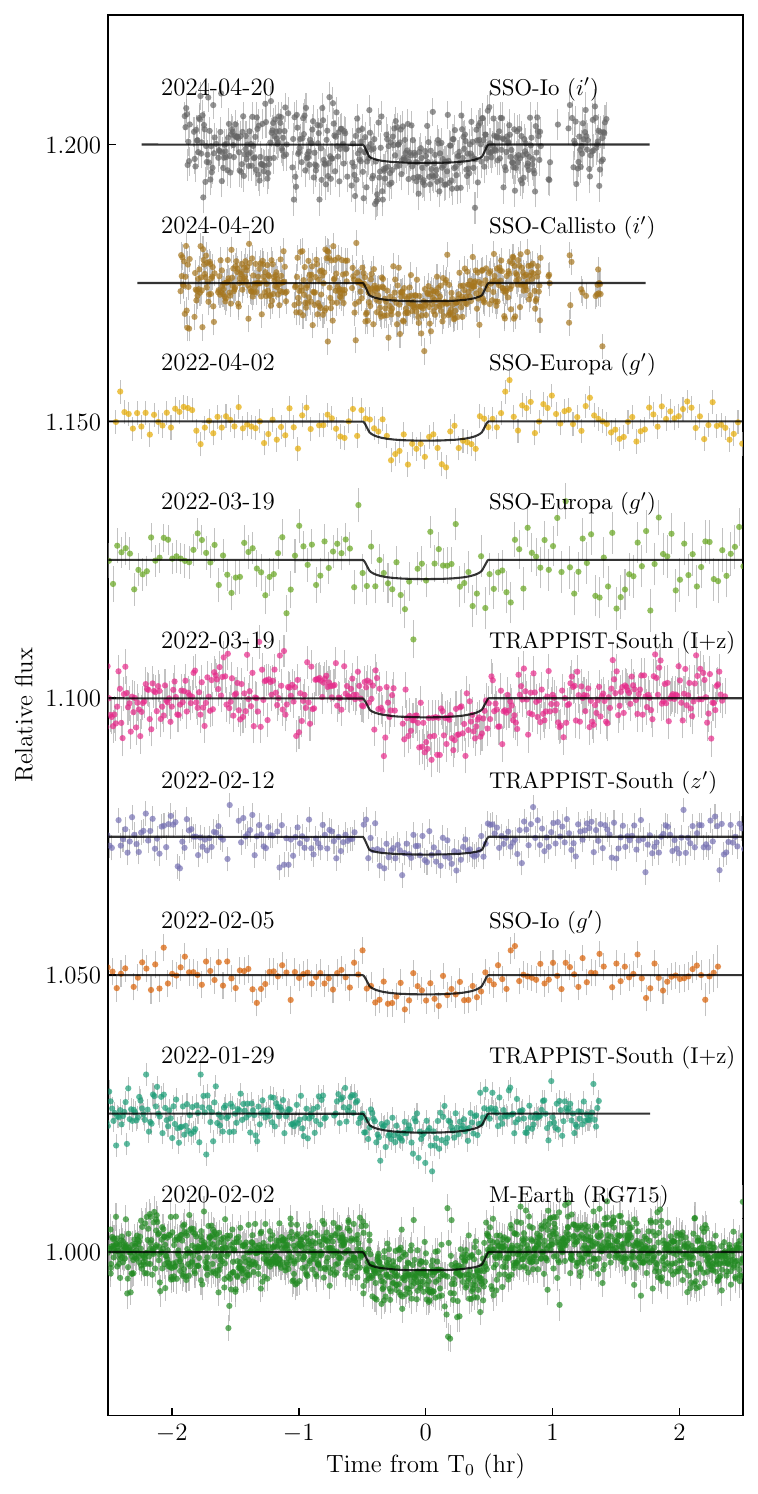}
  \caption{Ground-based photometric light curves of TOI-771 b. The solid black line shows the best-fitting model.
  }
    \label{fig:ground_based}
\end{figure}

\begin{figure}
\centering
  \includegraphics[width=\linewidth]{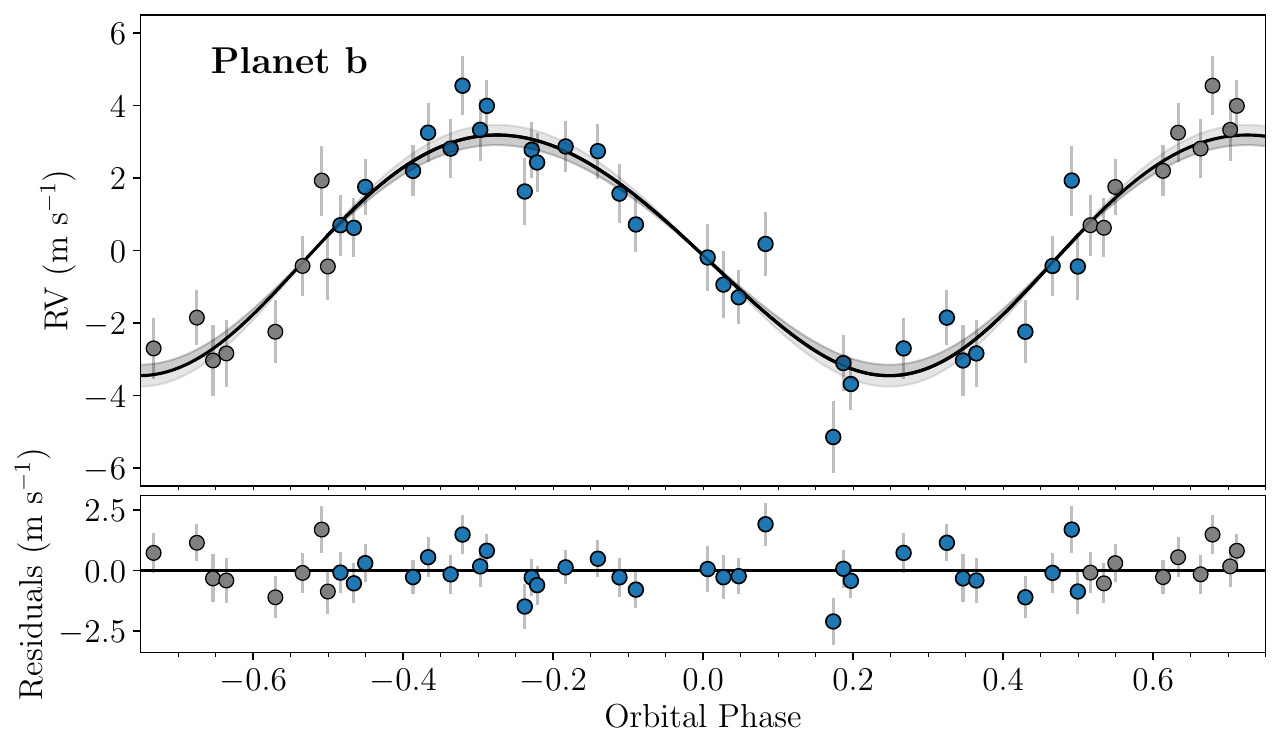}
  \includegraphics[width=\linewidth]{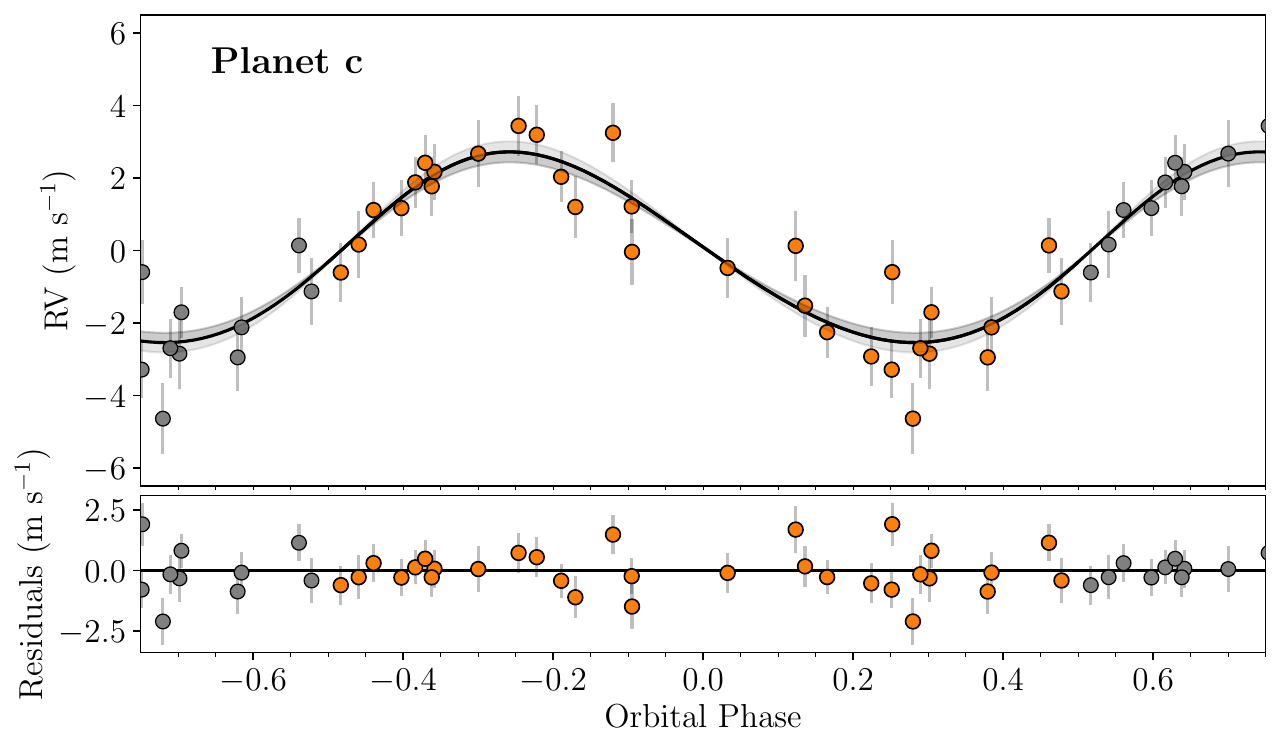}
  \caption{Phase-folded RVs of TOI-771 b and c and global RV model from the joint analysis. In the phase-folded plots, the coloured points shows the ESPRESSO data (with over-phased points in grey), and the solid black line shows the best-fitting model, with the $\pm 1 \sigma$ region displayed as a shaded area. Residuals are shown in the bottom panel. The jitter term has been added in quadrature to each error bar.
  }
    \label{fig:RV_phase}
\end{figure}

\begin{figure}
\centering
  \includegraphics[width=\linewidth]{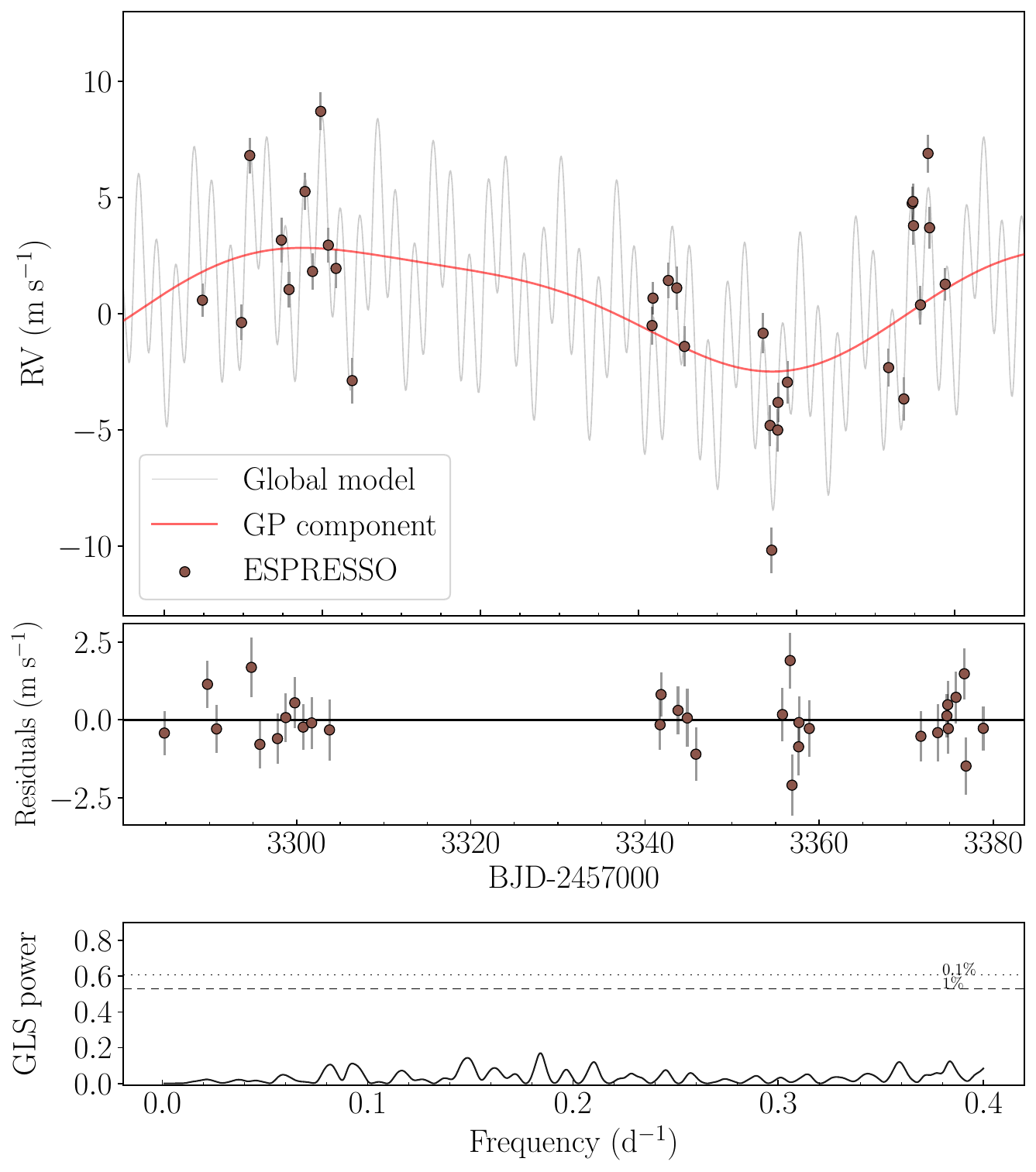}
  \caption{TOI-771 RV model from the joint fit. The solid grey line shows the best-fit global model, while the solid red line refers to GP component. Residuals are shown in the central panel. The jitter term has been added in quadrature to each error bar. The GLS periodogram of the residuals is shown in the bottom panel. The $1$\% and $0.1$\% FAP levels are marked with a horizontal dashed and dotted line, respectively. No significant signals are identified in the periodogram.
  }
    \label{fig:RV_global}
\end{figure}

We analysed all the \textit{TESS} and ground-based light curves together with the spectroscopic data using \pyorbit, adopting the two-planet model as described in Sect.~\ref{sec:RV_fit_only} and the photometric parameters as described in Sect.~\ref{sec:photometric_fit_only}.
In the joint fit, we also included the $S$-index time series as a proxy of the stellar activity, due to the significant signal detected around the expected rotational period of the star (Sect.~\ref{sec:star_activity}), and we modelled it together with the RVs, to better inform the GP \citep{Osborn2021, langellier2021,rajpaul2021, barragan2023}.
We assumed \prot, $P_\mathrm{dec}$, and {\it w} as common GP hyper-parameters for the two covariance matrices, while we independently fitted the amplitude of each covariance matrix.
For the light curve modelling, we fitted the LD coefficients independently for each filter, using the \texttt{PyLDTk}\footnote{\url{https://github.com/hpparvi/ldtk}.} \citep{Husser2013, Parviainen2015} estimates as initial values with a $0.1$ uncertainty as Gaussian priors. 
We adopted the \textit{TESS} model described in Sect.~\ref{sec:photometric_fit_only}, while for each ground-based dataset we included a quadratic term in the modelling to account for potential systematics, and an additional jitter term.
Table~\ref{table:joint_parameters} reports our best-fitting parameters.
Figures \ref{fig:TESS_phase} and \ref{fig:ground_based} show the transit models from \textit{TESS} and ground-based photometry, respectively. The phase-folded RVs and the global RV model are displayed in Figs.~\ref{fig:RV_phase} and \ref{fig:RV_global}.

We derived precise planetary parameters for TOI-771 b, a super-Earth with $R_{\rm b} = 1.36 \pm 0.10$~\rearth, $M_{\rm b} = 2.47_{-0.31}^{+0.32}$~\mearth\ (from $K_{\rm b} = 3.29 \pm 0.28$~\ms), and a bulk density of $\rho_{\rm b} = 5.4 \pm 1.4$~\gcm. The planet orbits with a period of $2.32$~d, implying an equilibrium temperatures of \teq~$=543_{-34}^{+28}$~K \footnote{Temperature is computed as $T_{\rm eq} = T_\star \, \biggl(\dfrac{R_\star}{2a}\biggr)^{1/2} \, [f(1-A_{\rm B})]^{1/4}$, assuming a null Bond albedo ($A_{\rm B} = 0$) and an efficient day-night heat redistribution efficiency ($f=1$, corresponding to $\varepsilon = 1$, as defined in \citealt{cowan2011}). However, close-in planets could have higher dayside temperature, implying stronger emission. In the case of no circulation ($f=8/3$,$\varepsilon = 0$), the dayside temperature of planet b would be $692_{-43}^{+37}$~K.}.
Additionally, we recovered a second periodic signal in the RVs, which we identify as the non-transiting planet TOI-771 c (see also Sect.~\ref{sec:planet_properties_c}), with a minimum mass of $M_{\rm c} = 2.87_{-0.38}^{+0.41}$~\mearth\  (from $K_{\rm c} = 2.58_{-0.25}^{+0.27}$~\ms) and a period of $7.61 \pm 0.03$~d. 
 
We also recover a broad value for the stellar rotational period at $P_{\rm rot} = 98_{-5}^{+10}$ days from our GP joint modelling of the RVs and $S$-index (see Fig.~\ref{fig:RV_global}), a periodicity that we also tentatively recovered in the ASAS-SN photometry (Sect.~\ref{sec:star_activity}). 
However, given the short baseline of the RV observations, a longer time-span is needed to definitely confirm the value of the stellar rotational period, which showed a couple of aliases in the RV dataset.

No additional signals have been identified in the ESPRESSO residuals after the joint fit (Fig.~\ref{fig:RV_global}). 
As a further confirmation, we performed an additional fit including a third Keplerian signal, from which we obtained consistent results for planets b and c, while the parameters of the additional signal did not converge. 
We therefore adopt the results described in this section as our final results, as no additional planets can be identified in the current RV and photometric (see Sect.~\ref{sec:sherlock}) datasets.

Finally, given the presence of an additional planet in the system, we investigated the presence of possible transit timing variations (TTVs) of TOI-771 b due to the gravitational interaction among the two planets.
We performed a photometric fit of both \textit{TESS} and ground-based data with \texttt{PyORBIT} as described in Section~\ref{sec:photometric_fit_only}, but leaving each T$_0$ as a free parameter. 
We then compared the observed transit times with the predicted ones derived from the linear ephemeris (T$_0$ and $P$ as in Table~\ref{table:joint_parameters}) to evaluate the TTVs presence. 
We could not identify any significant TTV trend (see Fig.~\ref{fig:TTV}), with all transit times being consistent with the linear ephemeris within $2 \sigma$. 
This is not surprising, giving that the period commensurability of the two planets is not close to any first-order mean motion resonance (see also Sect.~\ref{sec:dynamical_analysis}), when the TTV signal is expected to be enhanced \citep{agol2005, holman2005}.

\begin{table}
\caption{Best-fitting parameters of the TOI-771 system.}
\label{table:joint_parameters} 
\small
\begin{threeparttable}[t]
\centering
\begin{tabular}{l c c} 
  \hline\hline       
 \multicolumn{3}{c}{Planetary parameters}\\[1ex]
 \hline 
   & TOI-771 b & TOI-771 c \\
  \hline
  % \multicolumn{3}{c}{Fitted parameters}\\[1ex]
  $P$ (d) & $2.3260155_{-0.0000009}^{+0.0000010}$ & $7.61 \pm 0.03$ \\
  $T_0$ (TBJD)$^a$  & $1572.4191_{-0.0003}^{+0.0003}$ & $1573.8 \pm 2.5$\\
  $a/$\rstar & $18.9 \pm  1.1$  & $42 \pm 2$  \\
  $a$ (AU)   & $0.0208_{-0.0010}^{+0.0009}$ & $0.046\pm 0.002$ \\
  $R_\mathrm{p}/$\rstar & $0.05369 \pm 0.0008$ & - \\
  \rplanet\ (\rearth) & $1.36 \pm 0.10$  & - \\
  $b$ & $0.22_{-0.16}^{+0.18}$ & - \\
  $i$ (deg) & $89.3 \pm 0.5$ & - \\
  $T_{14}$ (h) & $0.96_{-0.05}^{+0.04}$ & - \\
  $e$ & $0.055_{-0.036}^{+0.047}$ & $0.065_{-0.045}^{+0.062}$ \\
  $\omega$ (deg)   & $-130_{-66}^{+38}$  & $-82_{-84}^{+88}$ \\
  $K$ (\ms)  & $3.29 \pm 0.28$ &$2.58_{-0.25}^{+0.27}$\\
  \mplanet\ (\mearth) & $2.47_{-0.31}^{+0.32}$  &-\\
  \mplanet\ $\sin{i}$\ (\mearth) & -  &$2.87_{-0.38}^{+0.41}$\\
  \rhoplanet\ (\gcm)  & $5.4 \pm 1.4$ & -\\
  $\rho / \rho_{\oplus , s}^{b}$  & $0.84 \pm 0.22$ & -\\
  $S_{\rm p}$ ($S_{\oplus}$)  & $14_{-3}^{+4}$ & $3.0_{-0.7}^{+0.6}$\\
  $T_{\rm eq}^c$ (K)  & $543_{-34}^{+28}$ & $365_{-19}^{+22}$\\
  $g_{\rm p}^d$ (m s$^{-2}$)  & $13 \pm 2$ & - \\[1ex]
 \hline 
 \multicolumn{3}{c}{Common parameters}\\
 \hline
  \rhostar\ (\rhosun) & \multicolumn{2}{c}{$16.8_{-2.7}^{+3.0}$} \\
  $u_{1, \mathrm{TESS}}$ & \multicolumn{2}{c}{$0.17 \pm 0.08$} \\
  $u_{2, \mathrm{TESS}}$ & \multicolumn{2}{c}{$0.40_{-0.08}^{+0.10}$} \\
  $u_{1, \mathrm{RG715}}$ & \multicolumn{2}{c}{$0.25 \pm 0.10$} \\
  $u_{2, \mathrm{RG715}}$ & \multicolumn{2}{c}{$0.31 \pm 0.10$} \\
  $u_{1, \mathrm{g'}}$ & \multicolumn{2}{c}{$0.26 \pm 0.09$} \\
  $u_{2, \mathrm{g'}}$ & \multicolumn{2}{c}{$0.54_{-0.10}^{+0.09}$} \\
  $u_{1, z'}$ & \multicolumn{2}{c}{$0.19 \pm 0.09$} \\
  $u_{2, z'}$ & \multicolumn{2}{c}{$0.27 \pm 0.10$} \\
  $u_{1, i'}$ & \multicolumn{2}{c}{$0.23 \pm 0.10$} \\
  $u_{2, i'}$ & \multicolumn{2}{c}{$0.36 \pm 0.09$} \\
  $u_{1, \mathrm{I + z}}$ & \multicolumn{2}{c}{$0.33 \pm 0.10$} \\
  $u_{2, \mathrm{I + z}}$ & \multicolumn{2}{c}{$0.37 \pm 0.10$} \\
  $\sigma_{\rm j, ESPRESSO}^{\rm e}$ (\ms) & \multicolumn{2}{c}{$0.72_{-0.24}^{+0.27}$} \\
  $\gamma_{\rm ESPRESSO}^{\rm f}$ (\ms) & \multicolumn{2}{c}{$-4.6 \pm 4.2$} \\[1ex]
   \hline 
 \multicolumn{3}{c}{GP hyper-parameters}\\
 \hline
  \prot\ (d) & \multicolumn{2}{c}{$98_{-5}^{+10}$} \\
  $P_{\rm dec}$ (\ms) & \multicolumn{2}{c}{$771_{-288}^{+169}$} \\
  $w$ (\ms) & \multicolumn{2}{c}{$0.95_{-0.07}^{+0.04}$} \\
  $A^{\rm g}_{\rm RV}$ (\ms) & \multicolumn{2}{c}{$4.9_{-1.6}^{+2.9}$} \\
  $A^{\rm g}_{\rm S-index}$ (\ms) & \multicolumn{2}{c}{$0.68_{-0.18}^{+0.34}$}\\[1ex]
 \hline
\end{tabular}
\tablefoot{
    \tablefoottext{a}{\tess\ barycentric Julian date (BJD$-2457000$).} 
\tablefoottext{b}{Scaled Earth’s bulk density, as defined in \cite{luquepalle2022}.} \tablefoottext{c}{$T_{\rm eq} = T_\star \, \biggl(\dfrac{R_\star}{2a}\biggr)^{1/2} \, [f(1-A_{\rm B})]^{1/4}$, assuming $f=1$ and $A_{\rm B} = 0$.}  \tablefoottext{d}{Planetary surface gravity.} 
\tablefoottext{e}{RV jitter term.}
\tablefoottext{f}{RV offset term.} 
\tablefoottext{g}{GP amplitude.}}
\end{threeparttable}
\end{table}

\subsection{Photometric planet searches and detection limits}
\label{sec:sherlock}

\begin{figure}
\centering
  \includegraphics[width=\linewidth]{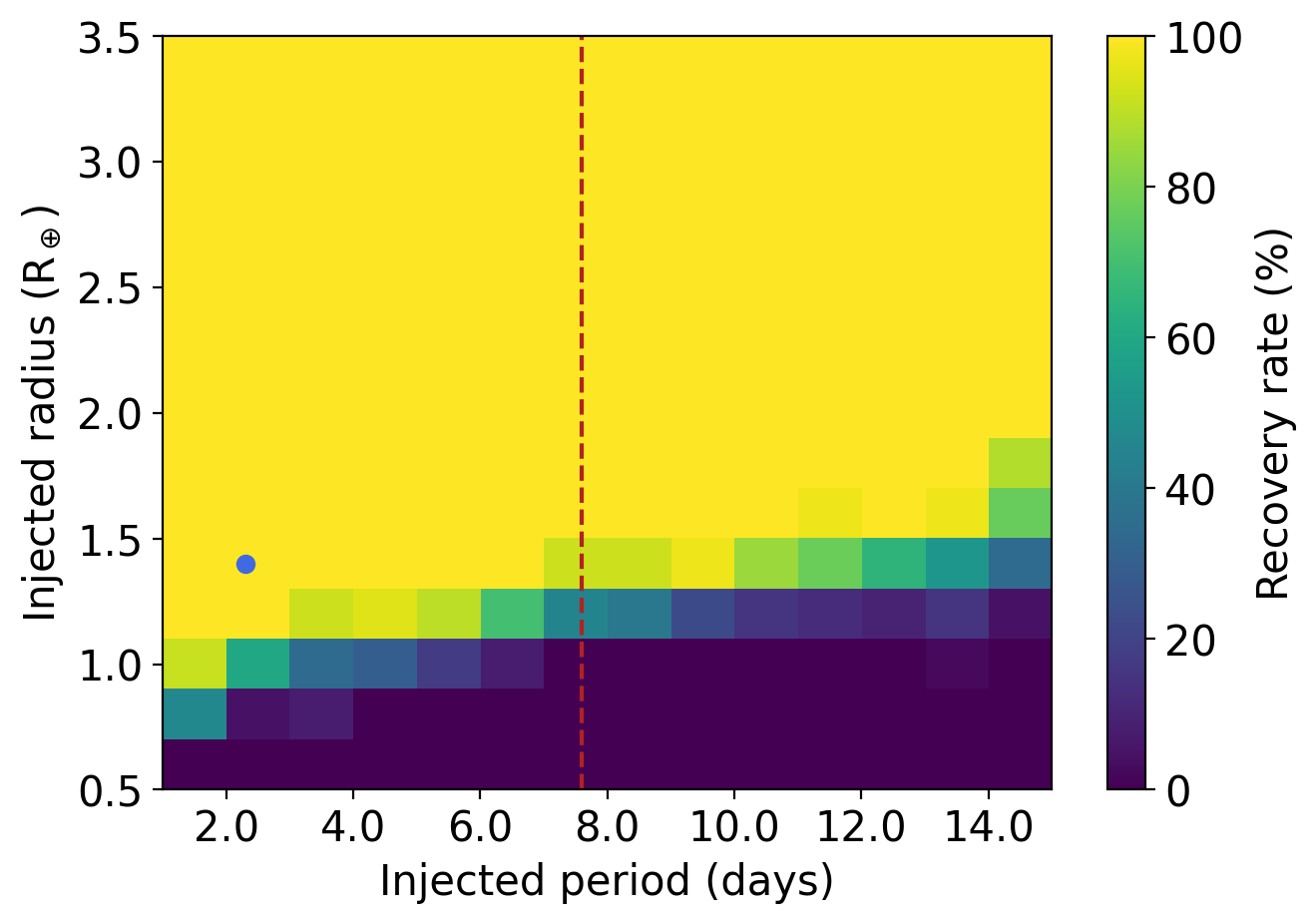}
  \caption{Detectability map of TOI-771. The injection-and-recovery experiment was conducted to establish the detection limits using the \textit{TESS} data. The blue dot marks the position of TOI-771\,b, while the dotted red line indicates the 7.6\,d period of TOI-771\,c as inferred from the RV fit. We explored 9000 different scenarios, where each pixel evaluates $\sim$45 of them. Larger recovery rates are presented in yellow and green colours, while lower recovery rates are shown in blue and darker hues.}
    \label{fig:inj_rec}
\end{figure}

As described in Sect.\ref{sec:global_analysis}, we found the presence of an extra planet in our RV dataset, with a period of $\sim 7.6$\,d. There is no alert associated with this orbital period coming from \textit{TESS}, which might hint at the non-transiting nature of 
this planet. Still, some transiting planets may remain undetected due to the high detection thresholds employed by official search pipelines, such as SPOC and Quick-Look Pipeline (QLP, \citealt{huang2020}), which effectively render low \snr\ ratio planets undetectable \citep[see, e.g.][]{delrez2022,peterson2023,gillon2024}. Hence, to confirm or rule out the transiting nature of this candidate, we carefully examined the available \textit{TESS} data using the SHERLOCK\footnote{\url{https://github.com/franpoz/SHERLOCK}} pipeline \citep{pozuelos2020,demory2020}. This pipeline is an open-source package that relies on six modules to (1) automatically acquire the data from an online database; (2) search for transiting planetary candidates; (3) perform a vetting of the detected signals; 
(4) conduct a statistical validation of the vetted signals; (5) refine the ephemerides by employing a Bayesian inference framework to model the transit light curves; and (6) find the upcoming transit windows observable from ground-based observatories. 
See \cite{pozuelos2023} for recent applications and different searching strategies, and \cite{devora_pajares2024} for further details of the most updated pipeline version.

In our first trial, we allowed SHERLOCK to search for signals with orbital periods ranging from 0.5 to 20\,d. After our first execution, we found a strong signal with a period of 2.32\,d, which corresponds to the known planet TOI-771\,b. During the subsequent runs, we found some additional weaker signals, all attributable to non-corrected noise or spurious results. We also conducted a dedicated search around the expected period of the planet TOI-771\,c, with a period grid ranging from 7.3 to 7.9\,d; however, we did not find any hint of a transiting signal in this range. 

The lack of a signal corresponding to the orbital period of TOI-771\,c does not necessarily confirm the non-transiting nature of this planet, as it could remain below our detection limit. To test this hypothesis, we conducted an injection-and-recovery experiment
that allows us to build a detectability map based on the \textit{TESS} data. To this end, we employed the MATRIX package\footnote{{The \texttt{MATRIX} (Multi-phAse Transits Recovery from Injected eXoplanets) code is open access on GitHub: \url{https://github.com/PlanetHunters/tkmatrix}}} \citep{devora2022}. MATRIX generates a sample of synthetic planets by combining a range of orbital periods, planetary radii, and orbital phases, it injects them into the \textit{TESS} light curve, and it executes a searching module mimicking the SHERLOCK algorithm. In particular, we generated 9000 scenarios by combining a grid of 30 periods (from 1 to 15\,d), 30 radii (from 0.5 to 3.5 $R_\oplus$), and 10 different epochs.   
Our results are displayed in Fig.~\ref{fig:inj_rec}, where we find that any transiting planet larger than 1.6\,R$_{\oplus}$ in our range of studied periods would have been detected using the current dataset, yielding recovery rates higher than 80$\%$. Hence, this result allows us to confirm the non-existence of any transiting planet with \rplanet~$\geq$1.6\,R$_{\oplus}$ with orbital periods shorter than 15\,d in the TOI-771 system. 
Considering the semi-amplitude derived from our RV fit (Sect.~\ref{sec:RV_fit_only}), and assuming the same inclination of TOI-771\,b ($i = 89.3 \pm 0.5$, see Sect.~\ref{sec:joint_fit}), 
we derived a minimum mass of $M_{\rm p} \sin{i} = 2.9 \pm 0.4$~\mearth\ for TOI-771\,c. Using \texttt{spright}\footnote{\url{https://github.com/hpparvi/spright}}, a Python Bayesian tool to compute the radius-density-mass relations for small planets \citep{parviainen2024}, we obtained a radius distribution with \rplanet\ lying between $1.3$~\rearth\ and $4.1$~\rearth\ (95\% limits), with the most probable value being \rplanet~$= 1.46$~\rearth\ (see also Sect.~\ref{sec:planet_properties_c}).
Our detectability map allows us to rule out most of the potential sizes for TOI-771\,c, and hence, the planet is most likely not transiting, even though there is a small probability of non-detection in the lower limit case of \rplanet~$= 1.3$~\rearth, where we have a recovery rate of $\sim$50$~\%$. In addition, our detectability map also allowed us to confirm that Earth and sub-Earth-sized planets are extremely challenging, if not impossible, to detect using \textit{TESS} photometry on such systems.

%--------------------------------------------------------------------
\section{Discussion}\label{sec:discussion}
\subsection{TOI-771 b planetary properties}\label{sec:planet_properties_b}

\begin{figure}
\centering
\includegraphics[width=\linewidth]{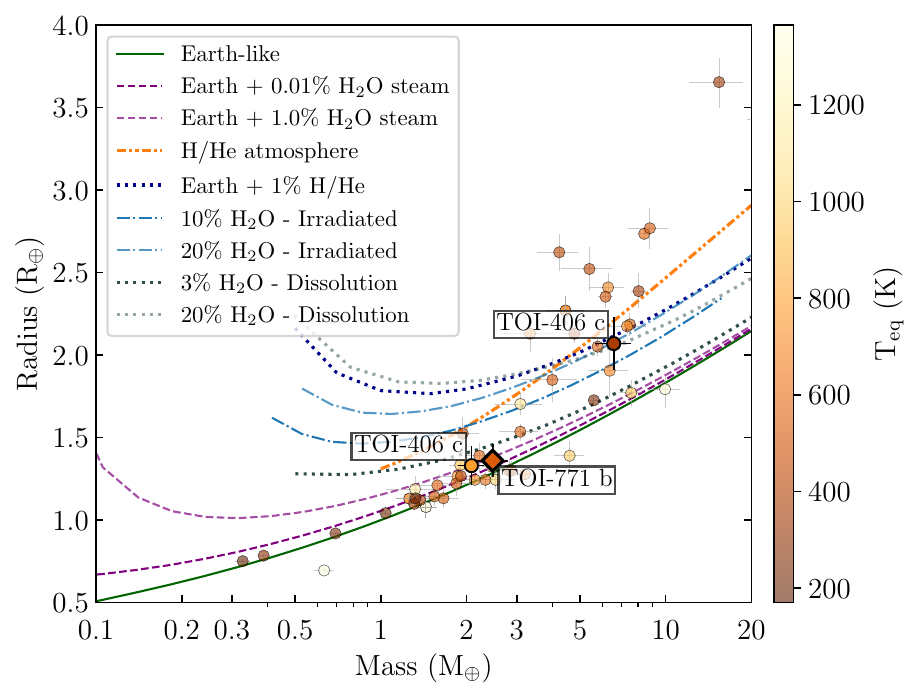}
  \caption{Mass-radius diagram for well-characterised M-dwarf exoplanets with \rplanet$< 4$~\rearth. The orange diamond marks the position of TOI-771 b. Previous \thirstee\ M-dwarf planets (TOI-406 b and c, \citealt{lacedelli2024}) are labelled and highlighted with solid black contours. Planets are colour-coded according to their equilibrium temperature. The planetary properties come from the PlanetS catalogue \citep{parc2024}, updated with the more recently published planets from the NASA Exoplanet Archive as of December 12, 2024, having mass and radius precision better than 25\% and 8\%, respectively. Different theoretical compositional tracks are show: \cite{Zeng2019} models for Earth-like composition (solid green line), and for Earth-like rocky cores with H/He atmospheres by different percentages in mass at $300$~K (dotted dark blue lines); \cite{Rogers23} mass-radius distribution from photoevaporation models (dotted orange line); \cite{turbet2020} models for water steam atmospheres (dotted purple lines); \cite{aguichine2021} models for irradiated ocean worlds at $500$~K with different water percentages (dash-dotted blue lines); and \cite{Luo24} models for rocky interior with water dissolution in both mantle and core with different water percentages (dotted grey lines).
  }
    \label{fig:MR_diagram}
\end{figure}

Following the \thirstee\ goal of identifying population patterns within sub-Neptunes, we compared the position of TOI-771 b in the mass-radius diagram with respect to the well-characterised population of planets with \rplanet~$< 4$~\rearth\ orbiting M dwarfs (Fig.~\ref{fig:MR_diagram}).
Very similarly to TOI-406 c, the first M-dwarf planet characterised within \thirstee\ \citep{lacedelli2024}, the density of TOI-771 b is slightly lower than Earth's, but still consistent with a rocky composition.
Considering its equilibrium temperature of $\sim 543$~K, higher than the runaway greenhouse limit \citep{turbet2020}, a supercritical steam atmosphere (that could be a consequence of magma ocean-hydrogen interactions \citealt{kite2020, Rogers2024}) could explain TOI-771 b's density, if water is present in a few percentages of mass fraction. 
Moreover, if water is present, the planet interior could be lighter also because of water solubility \citep{DornLichtenberg21, Luo24}.

\begin{figure*}
\centering
  \includegraphics[width=0.49\linewidth]{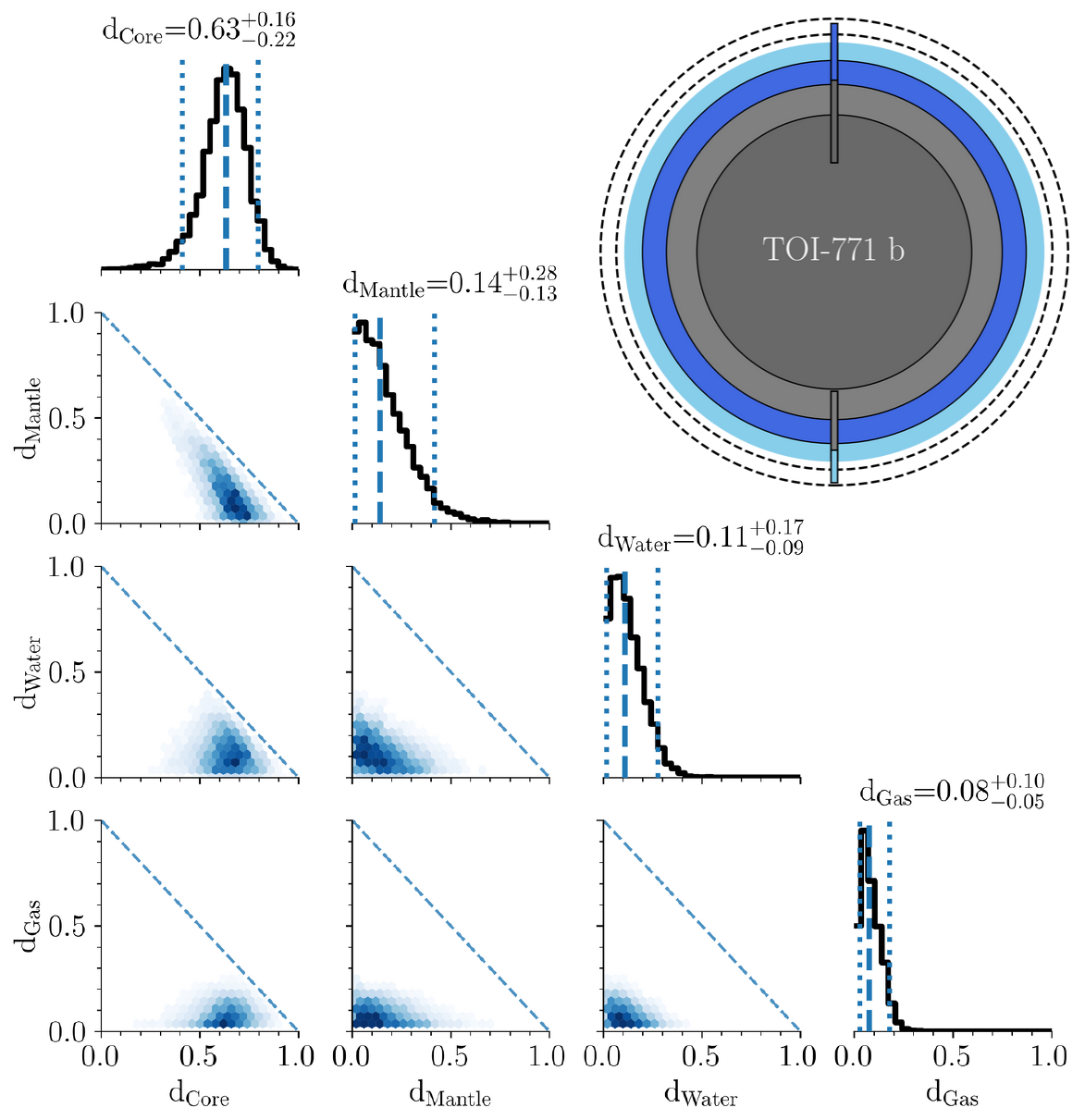}
  \includegraphics[width=0.49\linewidth]{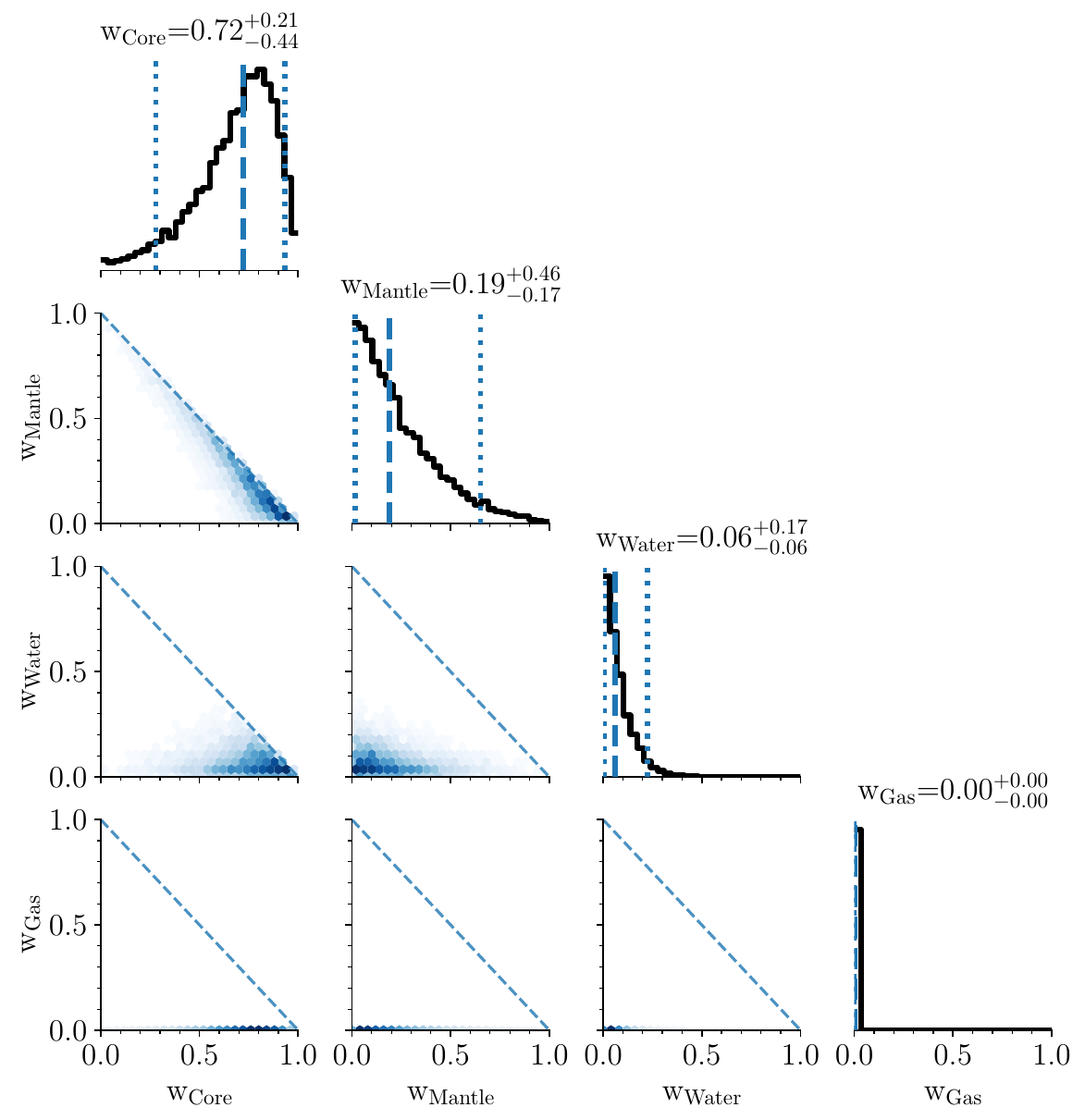}
  \caption{{\it Left panel}: Predicted thickness (radius fraction, d) of the interior layers of TOI-771 b as computed with \texttt{ExoMDN}. The panel also features a sketch of the planet showing the radius fractions corresponding to the posterior distributions of the inner core, mantle, water layer, and gas envelope in dark grey, light grey, dark blue, and light blue, respectively. The total uncertainty on the planetary radius is shown with the dashed black outer rings, while the uncertainties on each layers thickness are shown by coloured rectangles. {\it Right panel}: Predicted mass fraction (w) of the interior layers of TOI-771 b as computed with \texttt{ExoMDN}.
  }\label{fig:ExoMDN_b}
\end{figure*}

In the attempt to quantify the water mass fraction of the planet, we performed interior modelling using \texttt{ExoMDN}\footnote{\url{ https://github.com/philippbaumeister/ExoMDN}} \citep{Baumeister2023}, a machine-learning algorithm that uses the observed planetary radius \rplanet, mass \mplanet, and equilibrium temperature \teq\ to infer the internal planet composition assuming a four fixed-layers model (iron core, silicate mantle, water and high-pressure ice layer, H/He envelope). 
Figure~\ref{fig:ExoMDN_b} shows the thickness (i.e. the radius fraction) and the mass fraction of the interior layers of TOI-771 b, using as input our derived \mplanet, \rplanet, and \teq\ (Table~\ref{table:joint_parameters}). 
As expected, when assuming the \texttt{ExoMDN} model properties, TOI-771 b has a prominent planetary iron core, taking up to $63$\% of the planet size and $72$\% of its mass. For the other layers, the model predicts a possibly considerable mantle ($19$\% in mass fraction), a relatively small water layer ($\sim 6$\% in mass fraction), and a negligible amount of gas. 

When compared to the synthetic population models of \citep{venturini2024}\footnote{Simulation data for plots are available on \url{https://zenodo.org/records/10719523}.} for planets around M dwarfs, TOI-771 b's position is consistent with the population of pure rocky cores. 
Hints of formation inside the water ice line arise from the small amount of volatiles suggested from the slightly lower planetary density with respect to a pure rocky core.

\begin{figure}
\centering
  \includegraphics[width=\linewidth]{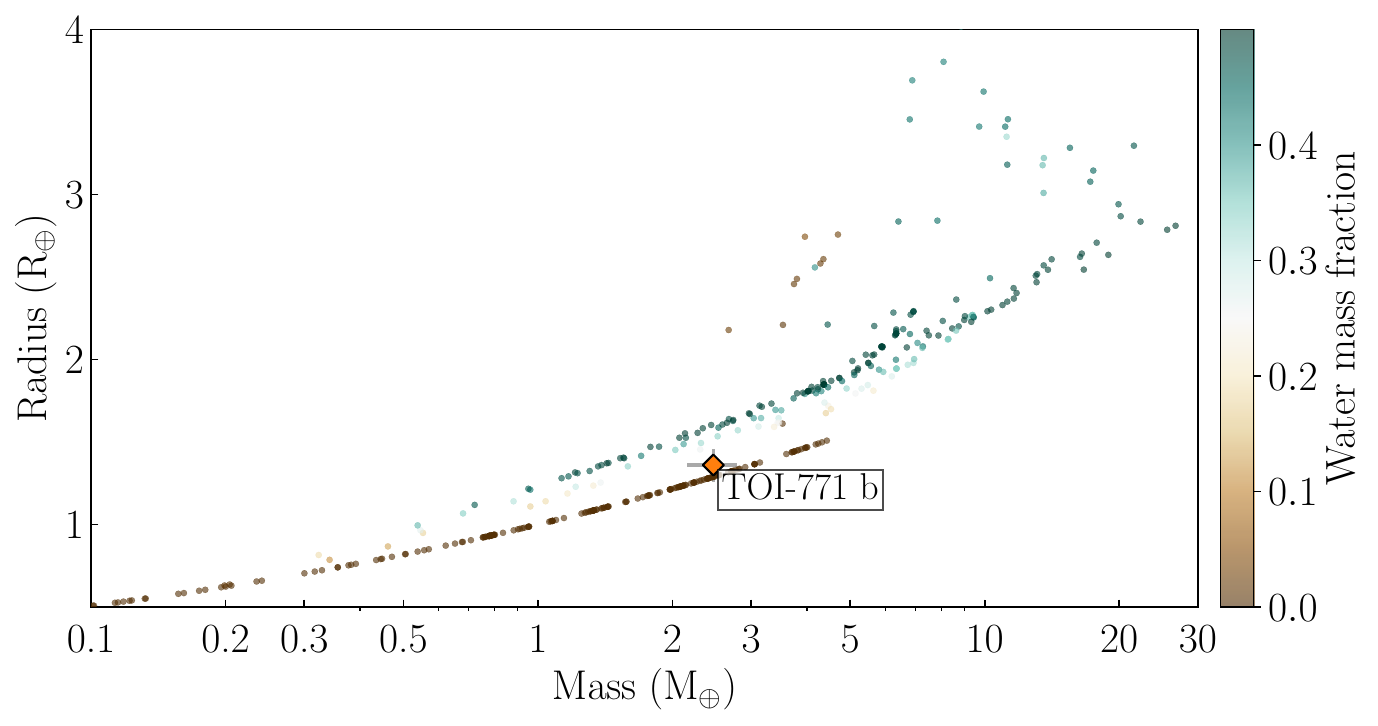}
  \caption{Theoretical mass-radius diagram for a synthetic population of planets at $2$~Gyr around an M dwarf with  \mstar~$=0.4$~\msun, orbiting with periods $P < 100$~d \citep{venturini2024}. Each point represents a synthetic planet, and it is colour-coded according to its water mass fraction, with brown colours indicating purely rocky cores, and blue colours showing planets with a substantial amount of water (up to $50$\%), as predicted from formation beyond the ice line \citep{alibert2013, raymond2018, bitsch2018, brugger2020, izidoro2021, Venturini20}. TOI-771 b is marked with an orange diamond, and it is consistent with the rocky population. 
  }\label{fig:MR_synthetic}
\end{figure}

In line with the previous \thirstee\ results, TOI-771 b adds to the population of planets around M dwarfs with bulk density comparable to Earth's identified in the $1.4-2$~\rearth\ regime by \cite{luquepalle2022}.
Such a population differs from a second, lower-density group of planets identified in the same radius range, which could be both enriched in volatiles \citep{luquepalle2022}, or it might host a gaseous envelope \citep{Rogers23}.
This result further supports the hypothesis of the density gap in the observed small exoplanet population orbiting M dwarfs, initially proposed by \cite{luquepalle2022}. 
The presence of such a density gap has been statistically confirmed by \cite{Schulze2024}, and theoretically supported by formation and evolution models \citep{venturini2024}, even though its significance and location is still debated, especially due to sample incompleteness in the low-mass regime \citep{parc2024}. 
A larger, well-characterised sample will be crucial to assess the statistical significance of this density gap, as generally aimed by the \thirstee\  project, and the characterisation of TOI-771 b significantly contributes to this goal.

\subsection{The non-transiting candidate TOI-771 c}\label{sec:planet_properties_c}
The retrieved semi-amplitude of TOI-771 c
% ($K_{\rm c} = 3.39 \pm 0.30$~\ms) 
implies a minimum mass of $M_{\rm c} \sin{i} = 2.87_{-0.38}^{+0.41}$~\mearth\ for the planet. We do not recover TOI-771 c with the current photometric dataset. Considering our injection-recovery tests, the planet is likely non-transiting, preventing the measure of the true planetary mass. 
For the TOI-771 system, even a small mutual inclination ($\Delta i \sim 1$~$\deg$) would allow planet c not to transit. In fact, considering the derived planetary and stellar parameters, the condition for showing a transit is $i_{\rm c} > 88.64$~$\deg$, where the inclination of the inner planet is $i_{\rm b} = 89.3 \pm 0.5$~$\deg$.

Even though inclination cannot directly be constrained, according to \cite{fisher2014} the planet has an 87\% probability of having an inclination between 30~$\deg $ and 90~$\deg$ \citep{fisher2014}, which implies a true mass lying within a factor two of the minimum mass, that is \mplanet$\lesssim 5.82$~\mearth. 
Assuming this mass value, the planet likely belongs to the sub-Neptune population \citep{parviainen2024}. 
Additional constraint on the inclination can be derived considering the presence of the additional transiting companion TOI-771 b. In fact, based on population statistical studies, multi-planet systems with $a/$\rstar$>5$ tend to have mutual inclination $\Delta i \lesssim 5$~$\deg$  \citep{tremaine2012, fang2012, Fabrycky14, Dai2018}.
Assuming a mutual inclination of $\Delta i = 5$~$\deg$ ($i_{\rm c} = 84.3$~$\deg$), the planet would have a mass of $M_{\rm c} = 2.92 \pm 0.42$~\mearth.
A similarity between the masses of the two planets (with $M_{\rm b} = 2.47_{-0.31}^{+0.32}$~\mearth), is also expected from the so-called `peas
in a pod' pattern, according to which planets in the same system are likely to have similar masses \citep{millholland2017, goyal2022}, as well as orbital spacings and radii \citep{weiss2018_peas}. 
Finally, from our dynamical and stability analysis (Sect.~\ref{sec:dynamical_analysis}), we derive a conservative mass upper limit of $11$~\mearth, which confirms the sub-Neptune nature of the planet.

With an equilibrium temperature of $365_{-19}^{+22}$~K, TOI-771 c belongs to the warm temperature regime (\teq$<400$~K), however it is located outside the empirical habitable zone of its star, which embraces planets with periods $12$~d~$<$~P~$<73$~d \citep{kopparapu2013}.  

\subsection{Dynamical analysis}
\label{sec:dynamical_analysis}
To investigate the possible proximity of the TOI-771 system to mean motion resonances (MMRs) and the dynamical stability of our best-fit solution, we used direct $N$-body integrations and a CPU-efficient fast indicator called the reversibility error method \citep[REM;][]{Panichi_2017}. 
REM is closely related to the maximum Lyapunov exponent (MLE). It is based on integrating the equations of motion with a time-reversible (symplectic) scheme back and forth for the same number of steps. Then the difference between the initial and final states of the system, normalised by the size of the phase-space trajectory, makes it possible to distinguish between regular and chaotic evolution. 
REM is scaled so that $\log\mathrm{\widehat{REM}}=0$ means that the difference between the initial and final states is of the order of the size of the orbit.

\begin{figure}[!h]
\centering
\includegraphics[width=0.925\linewidth]{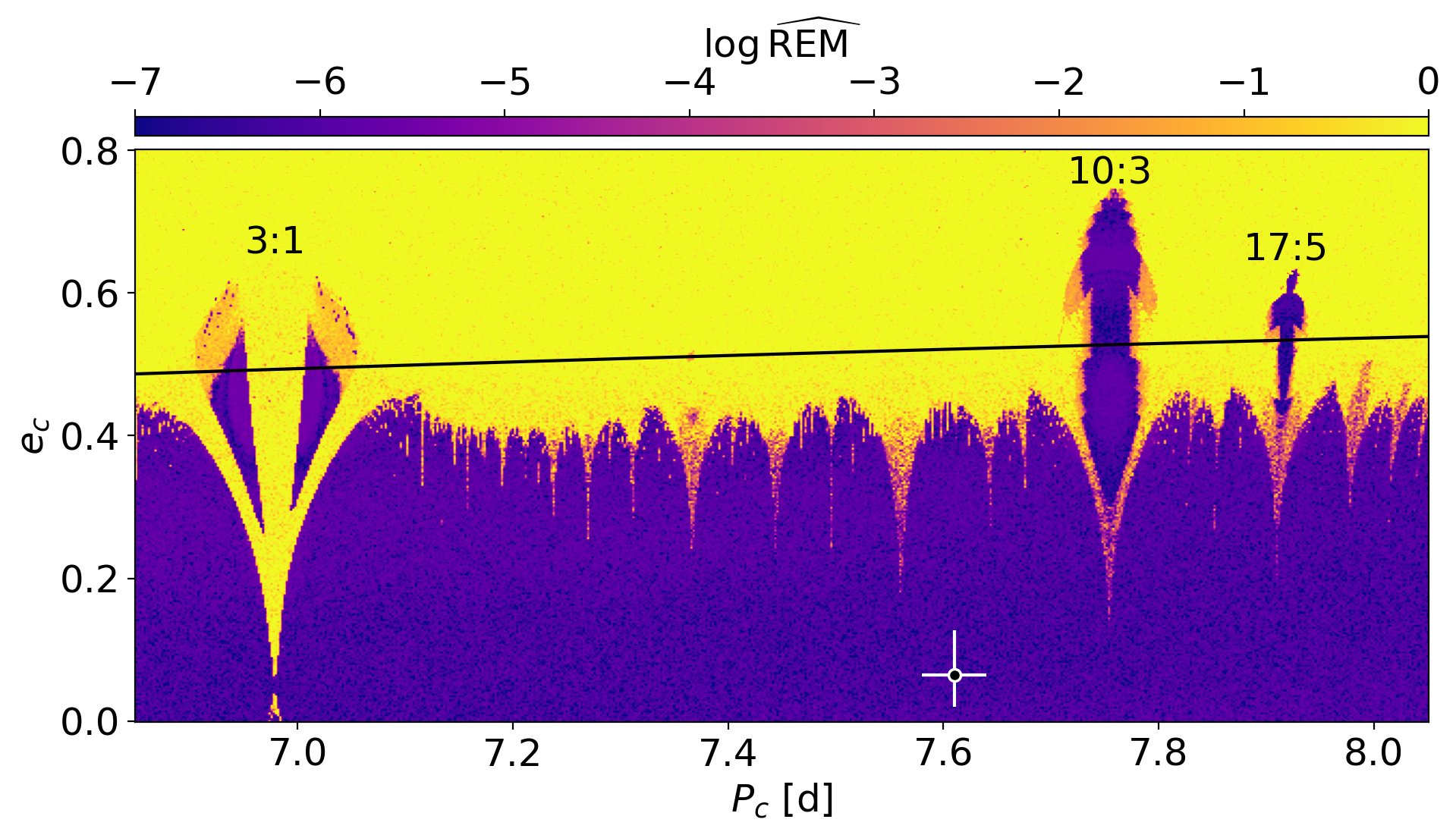}
\caption{
REM dynamical map for the solution presented in Table~\ref{table:joint_parameters} in the $(P_{\rm c},e_{\rm c})$ plane of planet c. Small values of $\log \mathrm{\widehat{REM}}$ characterise regular (long-term stable) solutions, marked with a dark blue colour. Dynamically chaotic solutions are marked with lighter colours, up to yellow. Some of the MMRs producing dynamical features in the map are highlighted. The black curve represents the geometric collision of the orbits, defined by the condition: $a_{\rm b}(1 + e_{\rm b}) = a_{\rm c}(1 - e_{\rm c})$. The white circle with crossed lines shows our best-fit solution with uncertainties. The resolution of the map is 601~$\times$~301 points.
}
\label{fig:mmrmap}
\end{figure}

We integrated REM with the \whfast{} integrator with the 17-th order corrector, with a fixed time step of $0.06$~d, as implemented in the \rebound{} package \citep{rein2012,Rein2015MNRAS.452..376R} for $\simeq 2\times 10^5$ orbital periods of the outer planet. 
As shown in Fig.~\ref{fig:mmrmap}, our best-fit solution has $\log\mathrm{\widehat{REM}}$ $\leq$ -5, indicating a stable solution. 
Furthermore, the location of the best-fit model in phase space indicates its non-resonant character. The dynamical map shows the nominal system shifted by a few $\sigma$ from the 10:3~MMR in $(P_{\rm c}, e_{\rm c})$ plane, and it shows a narrow limitation in the $3 \sigma$ range due to the high-order resonance structure.

In the attempt to constrain the mutual inclination of the orbits using dynamics, we first computed the REM  maps in the $(\Delta\Omega, i_{\rm c})$ plane, where $\Delta\Omega \in [0^{\circ}, 360^{\circ}]$ is the separation of the nodal lines of the orbits and $i_{\rm c} \in [1^{\circ}, 179^{\circ}]$ is the inclination of planet~c. Then, the mutual inclination of the orbits is:
\begin{equation}
i_{\rm mut} = \cos i_{\rm b} \cos i_{\rm c} + 
\sin i_{\rm b} \cos i_{\rm c} \cos\Delta\Omega . 
\end{equation}

\begin{figure}[!h]
\centering
\includegraphics[width=0.925\linewidth]{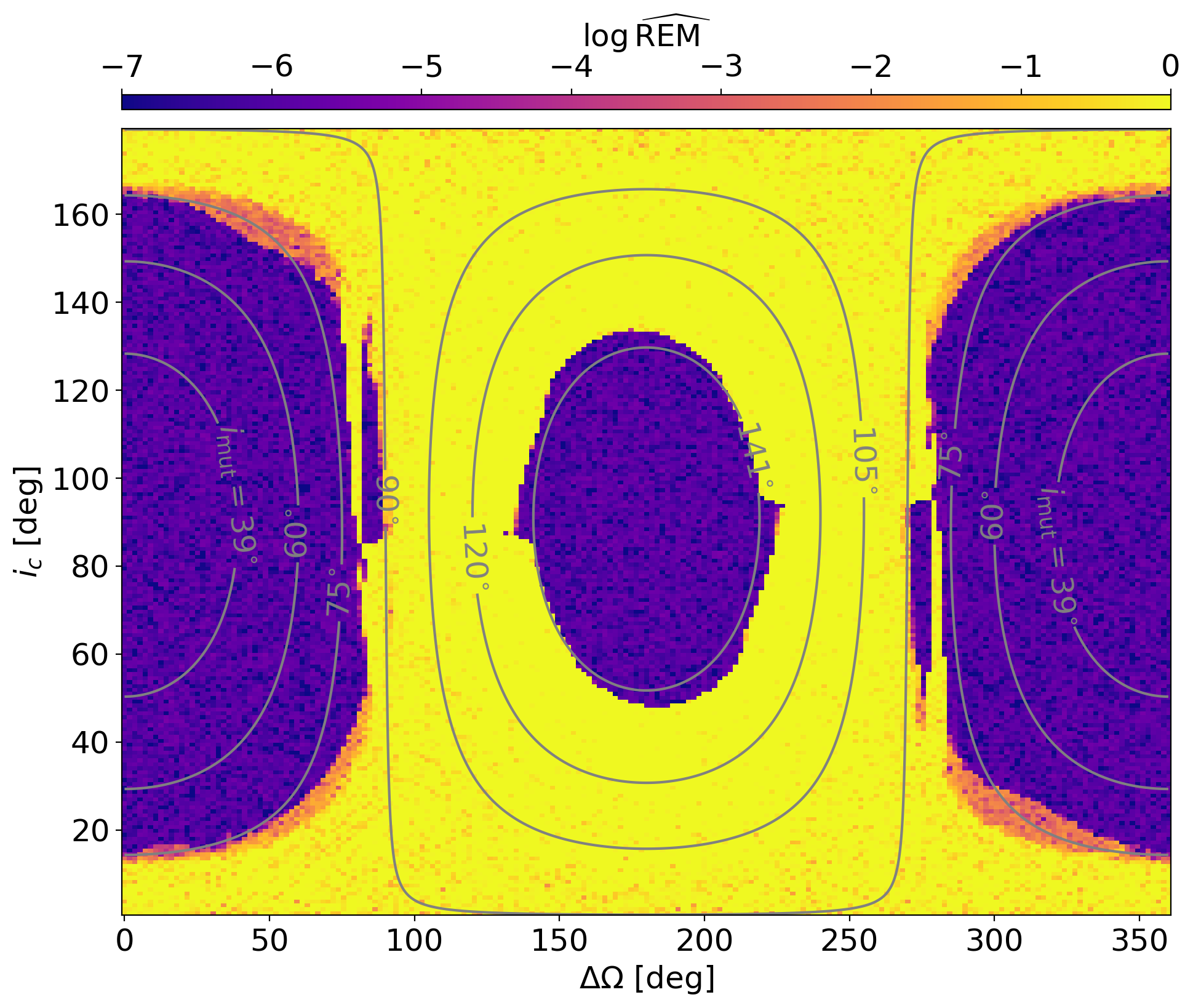}
\caption{Dynamic REM map for several cycles of oscillation of the elements in the $(\Delta\Omega, i_{\rm c})$ plane. In addition to the gravitational interactions, GR perturbations are also included. The grey curves mark the levels of mutual inclination of the orbits.}
\label{fig:mutual}
\end{figure}

We show in Fig.~\ref{fig:mutual} the results of the REM simulations, computed with an integration step of $0.02$~d.
Perhaps non-intuitively, stable regions appear only in a portion of the parameter plane, outside mutual inclinations of approximately $i_{\rm mut} \in (75^{\circ}, 140^{\circ})$. 
This implies moderate mutual inclinations, in line with the discussion presented in Sect.~\ref{sec:planet_properties_c}. Considering that $i_{\rm b} = 89.3^{\circ} \pm 0.5^{\circ}$, the inclination of TOI-771 c for a stable configuration is restricted to $i_{\rm c} \in (15^{\circ}, 165^{\circ})$, and therefore its mass must be at most $M_{\rm c} \simeq 11$~\mearth. 

The resulting scan of the $(\Delta\Omega, i_{\rm b})$ plane is reminiscent of the results for the Lidov-Kozai resonance \citep[e.g.][]{Shevchenko2017,Naoz2013}. Due to the fact that the LK resonance can be suppressed by general relativity (GR) perturbations (and star-planet tides) in the calculations we included the GR perturbations in the simplest, yet quantitatively correct and CPU-efficient form of the radial and conservative potential $\sim r^{-2}$ introduced by \cite{Nobili1986}. The interactions considered make it possible to consciously reduce the range of external body masses.

\subsection{Prospects for atmospheric characterisation}\label{sec:atmosphere}

\begin{figure*}[h]
\centering
  \includegraphics[width=0.49\linewidth]{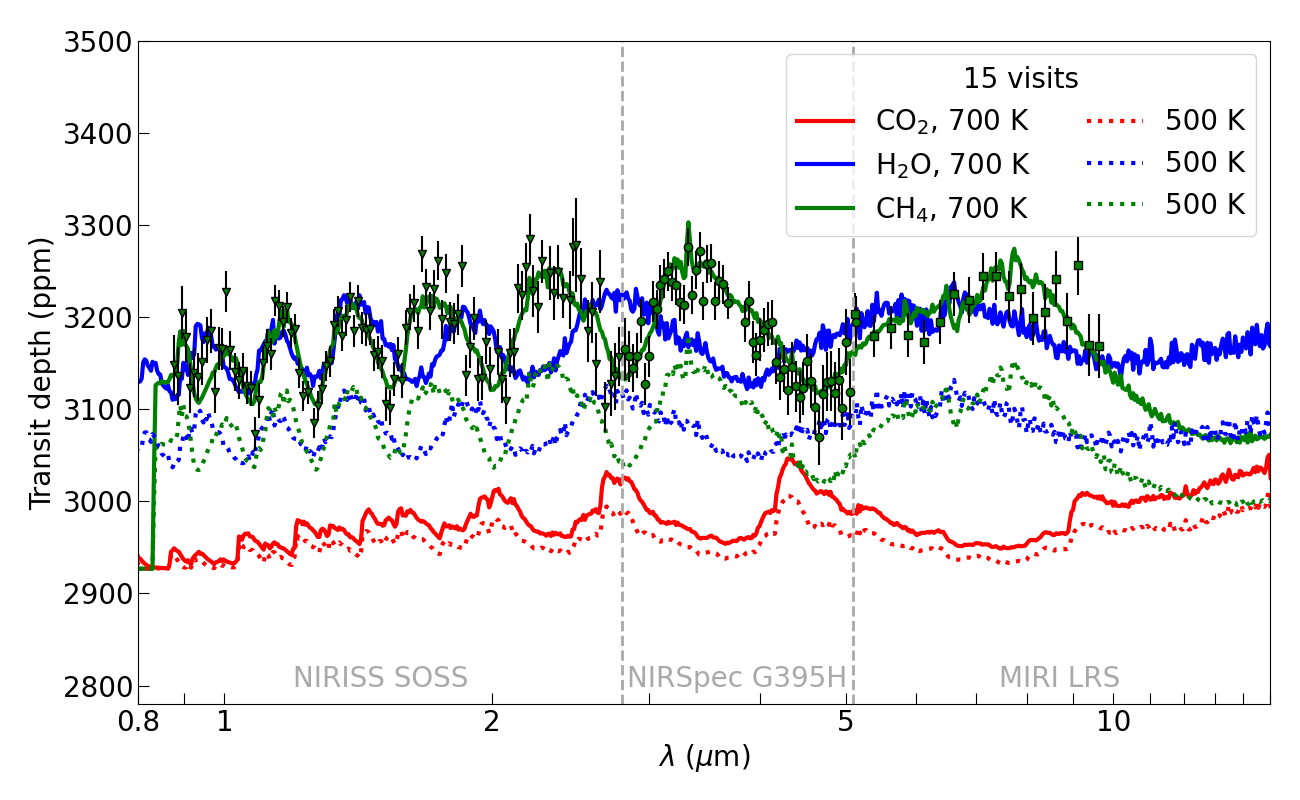}
  \includegraphics[width=0.49\linewidth]{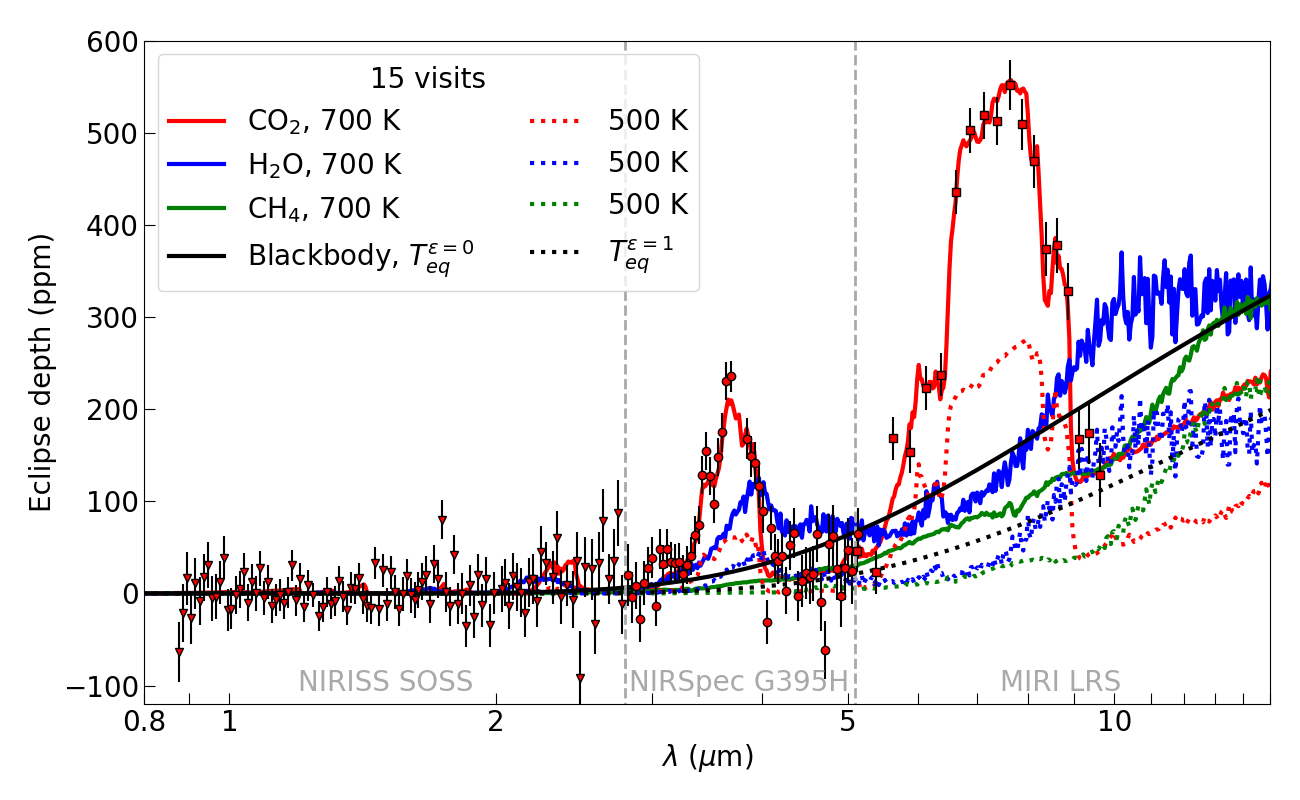}  
  \caption{Modelled transmission (left) and emission (right) spectra of TOI-771 b, assuming an atmospheric composition of 100\% CO$_2$ (red), 100\% H$_2$O (blue), and 100\% CH$_4$ (green) at different temperatures. For each composition we considered T-P profiles with equilibrium temperatures of 700\,K (continuos) and 500\,K (dashed line). Points with error-bars shows the simulated spectra for 15 transits (or eclipses) using the \textit{JWST} NIRISS-SOSS, NIRSpec G395H, and MIRI LRS configurations. }
    \label{fig:jwst_sim}
\end{figure*}

To investigate the feasibility of TOI-771 b's atmospheric characterisation, we calculated its transmission spectroscopy metric (TSM) and emission spectroscopy metric (ESM) following \cite{kempton2018}.
With TSM~$=15.5$ and ESM~$=4.7$,
TOI-771 b lies among the 
most favourable targets for both transmission and emission spectroscopy. In fact, the planet has a TSM higher than the threshold of $10$ proposed by \citealt{kempton2018} for \rplanet~$<1.5$~\rearth, and it is 
classified among the top five best-in-class targets for emission spectroscopy according to \citep{hord2024}.
The high precision on the planetary mass obtained from our analysis ($13$\%) makes it an even more appealing target. In fact, a mass precision of at least $20$\% is needed to break the degeneracy on compositional models and retrieve accurate atmospheric composition \citep{batalha19}.

We investigated the suitability of TOI-771 b for transmission and emission spectroscopy with the \textit{JWST} by performing spectral simulations based on a tailored set of atmospheric models.
We adopted \texttt{TauREx 3} \citep{al-refaie2021} to generate the synthetic spectra, using the stellar and planetary parameters from Tables \ref{table:star_params} and \ref{table:joint_parameters}. The atmospheric equilibrium temperature falls within the range of $\sim$500--700\,K, depending on its Bond albedo and circulation efficiency, as derived in Section \ref{sec:joint_fit}. We selected suitable temperature-pressure (T-P) profiles from the grid provided by \cite{Kempton_2023}.
TOI-771 b has most likely lost its primordial H/He envelope, but it may possess a secondary atmosphere formed by mantle outgassing.
We focused on three end-member compositions for the volatile-rich atmosphere: a water-world scenario with 100\% H$_2$O, a Venus-like atmosphere with 100\% CO$_2$, and an intriguing methane-world with 100\% CH$_4$. Additionally, for completeness, we considered H/He atmospheres with 1$\times$ and 100$\times$ scaled solar abundances under chemical equilibrium, in both clear and hazy conditions, similar to those described in previous papers (e.g. \citealp{Luque22,orell-miquel2023,goffo2024}).

We used \texttt{ExoTETHyS} \citep{morello2021} to simulate the corresponding \textit{JWST} spectra, as observed with the NIRISS-SOSS (0.6--2.8\,$\mu$m), NIRSpec-G395H (2.88--5.20\,$\mu$m), and MIRI-LRS (5--12\,$\mu$m) instrumental modes. We adopted a spectral resolution of $R\sim$100 for NIRISS-SOSS and NIRSpec-G395H, and a constant bin size of 0.25\,$\mu$m for MIRI-LRS, following the recommendations of the \textit{JWST} Transiting Exoplanet Community Early Release Science team \citep{carter2024,powell2024}.
The predicted error bars for a single transit observation are 49--197\,ppm (mean error 88 ppm) for NIRISS-SOSS, 66--139 ppm (mean error 98 ppm) for NIRSpec-G395H, and 88--134 ppm (mean error 105 ppm) for MIRI-LRS.

Figure \ref{fig:jwst_sim} shows the modelled transmission and emission spectra for the three steam atmosphere compositions, each paired with two distinct T-P profiles corresponding to equilibrium temperatures of 500\,K and 700\,K. The figure also overlays simulated spectra from 15 combined visits, for visualization purposes, for each instrumental mode.
The transmission spectra exhibit small features of $\sim$100 parts per million (ppm) for the methane and water-world scenarios, with even smaller features for the Venus-like composition. This behaviour is unsurprising, as CO$_2$ has the largest molecular weight, corresponding to smaller atmospheric scale heights. The T-P profiles also affect the atmospheric scale heights, but the wavelength dependence is negligible, practically resulting in small offsets between models with the same chemistry.
The emission spectra may display larger features at wavelengths $>$3\,$\mu$m, reaching up to $\sim$600 ppm for the hotter CO$_2$ model and $\sim$300 ppm for the hotter H$_2$O model. The methane-world emission is the most similar to a blackbody spectrum. For all cases in emission, the spectral amplitudes are roughly halved when the cooler T-P profiles are assumed.

Observing multiple secondary eclipses of TOI-771 b with MIRI-LRS and/or NIRSpec-G395H would provide the best opportunity to determine its atmospheric composition. In the most favourable case of a Venus-like atmosphere with low circulation efficiency (dayside temperature $\sim$700 K), even one or two eclipses may suffice to detect strong CO$_2$ features. However, additional observations would be needed to distinguish a water-world from a methane-world or other configurations leading to blackbody-like spectra.

Observing multiple primary transits of TOI-771 b with NIRISS-SOSS and/or NIRSpec-G395H offers a viable alternative to detect an H$_2$O or CH$_4$-dominated atmosphere, but is less effective for heavier molecular weight compositions such as those of Venus and Earth. In the unlikely case of an H/He-dominated envelope with up to $\sim$100$\times$ solar metallicity, clear or hazy conditions, a single transit observation could easily reveal molecular absorption features (models not shown). However, clouds may attenuate absorption features in transmission, hindering their detection regardless of the atmospheric chemical composition. Another challenge for transit spectroscopy may arise from unocculted stellar spots on the M dwarf host, which can mimic water-world absorption features (e.g. \citealp{moran2023}).

In summary, our simulations identify eclipse spectroscopy as the most promising technique to robustly detect and characterise the atmosphere of TOI-771 b (if present), owing to the possibility of larger spectral features in the mid-IR and a lower risk of contamination from stellar activity compared to transit spectroscopy. The success of this approach also depends on the atmospheric composition. For instance, a methane-world would emit less than 100 ppm at wavelengths below 10 $\mu$m, with a spectrum closely resembling that of an airless, reflective blackbody at the planetary equilibrium temperature. These scenarios could be distinguished by incorporating a precise measurement with the MIRI 15 $\mu$m photometer.

%--------------------------------------------------------------------
\section{Conclusions}\label{sec:conclusions}

In this work, we used \textit{TESS} and ground-based photometry, together with ultra-precise ESPRESSO RVs, to analyse one of the targets of the \thirstee\ project, the M-dwarf star TOI-771. 
We derived precise planetary parameters for TOI-771 b, a super Earth with an orbit of $2.32$~d, and radius and mass of $R_{\rm b} = 1.36 \pm 0.10$~\rearth, $M_{\rm b} = 2.47_{-0.31}^{+0.32}$~\mearth.
Its bulk density ($\rho_{\rm b} = 5.4 \pm 1.4$~\gcm) makes it consistent with a rocky composition, even though a small amount of volatiles could be present. 
Additionally, we inferred from the RV dataset the presence of a second, non-transiting sub-Neptune in the system, TOI-771 c, with a minimum mass of $M_{\rm c} \sin{i} = 2.87_{-0.38}^{+0.41}$~\mearth\  and a period of $7.61 \pm 0.03$~d.

In line with previous \thirstee\ results, TOI-771 b supports the evidence of two populations with different composition around M dwarfs, separated by a density gap. 
With a high-precision mass ($13$\%) and a warm effective temperature of $543_{-34}^{+28}$~K, the planet is a particularly interesting target for atmospheric characterisation, and it is included in the list of targets under consideration of the Rocky World Director's Discretionary Time (DDT) programme\footnote{\url{https://outerspace.stsci.edu/display/HPR/The+Rocky+Worlds+DDT+Program\%3A+Implementation\%2C+Structure+and+Policies}}, a 500-hour \textit{JWST} and \textit{HST} programme to investigate the atmosphere of terrestrial exoplanets orbiting M dwarfs \citep{redfield24}.
Our simulations show that \textit{JWST} eclipse spectroscopy is the most effective approach to detect and characterise the atmosphere of this planet (if present), and distinguish between a methane, CO$_2$, or water-dominated atmospheric composition.

The characterisation of the TOI-771 system provides a valuable addition to the sample of small, well-characterised planets around M dwarfs, pursuing one of the goals of the \thirstee\ project by enlarging the sample of well-characterised planets across spectral types, with the ultimate aim of understanding the sub-Neptune population under a global perspective.

\section*{Data availability}\label{sec:data_availability}

The ESPRESSO RVs and activity indicators (Table \ref{table:RV_table_ESPRESSO}) are available in electronic form at the CDS via anonymous ftp to cdsarc.u-strasbg.fr (130.79.128.5) or via \url{http://cdsweb.u-strasbg.fr/cgi-bin/qcat?J/A+A/}.

\begin{acknowledgements}
We acknowledge financial support from the Agencia Estatal de Investigaci\'on of the Ministerio de Ciencia e Innovaci\'on MCIN/AEI/10.13039/501100011033 and the ERDF “A way of making Europe” through project PID2021-125627OB-C32, and from the Centre of Excellence “Severo Ochoa” award to the Instituto de Astrofisica de Canarias.
This paper includes data collected by the {\it TESS} mission,
which are publicly available from the Mikulski Archive for Space
Telescopes (MAST). Funding for the {\it TESS} mission is provided 
by the NASA Explorer Program. 
Resources supporting this work were provided by the NASA High-End Computing (HEC) Program through the NASA Advanced Supercomputing (NAS) Division at Ames Research Center for the production of the SPOC data products.
This research has made use of the Exoplanet Follow-up Observation Program (ExoFOP; DOI: 10.26134/ExoFOP5) website, which is operated by the California Institute of Technology, under contract with the National Aeronautics and Space Administration under the Exoplanet Exploration Program. 
Funding for the TESS mission is provided by NASA's Science Mission Directorate. 
This research has made extensive use of the SIMBAD
database, operated at CDS, Strasbourg, France, and NASA’s Astrophysics Data System.
This research has
made use of the NASA Exoplanet Archive, which is
operated by the California Institute of Technology, 
under contract with the National Aeronautics and Space
Administration under the Exoplanet Exploration Programme. 
This work has made use of data from the European Space Agency (ESA) mission
{\it Gaia} (\url{https://www.cosmos.esa.int/gaia}), processed by the {\it Gaia}
Data Processing and Analysis Consortium (DPAC,
\url{https://www.cosmos.esa.int/web/gaia/dpac/consortium}). 
Funding for the DPAC
has been provided by national institutions, in particular the institutions
participating in the {\it Gaia} Multilateral Agreement.
This publication makes use of data products from the Two
Micron All Sky Survey, which is a joint project of the
University of Massachusetts and the Infrared Processing and
Analysis Center/California Institute of Technology, funded by
the National Aeronautics and Space Administration and the
National Science Foundation. 
This work makes use of observations from the LCOGT network. Part of the LCOGT telescope time was granted by NOIRLab through the Mid-Scale Innovations Programme (MSIP). MSIP is funded by NSF.
Based on data collected by the SPECULOOS-South Observatory at the ESO Paranal Observatory in Chile. The ULiege's contribution to SPECULOOS has received funding from the European Research Council under the European Union's Seventh Framework Programme (FP/2007-2013) (grant Agreement n$^\circ$ 336480/SPECULOOS), from the Balzan Prize and Francqui Foundations, from the Belgian Scientific Research Foundation (F.R.S.-FNRS; grant n$^\circ$ T.0109.20), from the University of Liege, and from the ARC grant for Concerted Research Actions financed by the Wallonia-Brussels Federation.
The Birmingham contribution is in part funded by the European Union's Horizon 2020 research and innovation programme (grant's agreement n$^{\circ}$ 803193/BEBOP), and from the Science and Technology Facilities Council (STFC; grant n$^\circ$ ST/S00193X/1, ST/W000385/1 and ST/Y001710/1).
The Cambridge contribution is supported by a grant from the Simons Foundation (PI Queloz, grant number 327127).
Based on data collected by the TRAPPIST-South telescope at the ESO La Silla Observatory. TRAPPIST is funded by the Belgian Fund for Scientific Research (Fond National de la Recherche Scientifique, FNRS) under the grant PDR T.0120.21, with the participation of the Swiss National Science Fundation (SNF). E.J. and M.G. are FNRS Research Directors.
This paper makes use of data from the MEarth Project, which is a collaboration between Harvard University and the Smithsonian Astrophysical Observatory. The MEarth Project acknowledges funding from the David and Lucile Packard Fellowship for Science and Engineering, the National Science Foundation under grants AST-0807690, AST-1109468, AST-1616624 and AST-1004488 (Alan T. Waterman Award), the National Aeronautics and Space Administration under Grant No. 80NSSC18K0476 issued through the XRP Programme, and the John Templeton Foundation.
This material is based upon work supported by the National Aeronautics and Space Administration under Agreement No.\ 80NSSC21K0593 for the programme ``Alien Earths''.
The results reported herein benefited from collaborations and/or information exchange within NASA's Nexus for Exoplanet System Science (NExSS) research coordination network sponsored by NASA's Science Mission Directorate.

Support for this work was provided by NASA through the NASA Hubble Fellowship grantcHST-HF2-51559.001-A awarded by the Space Telescope Science Institute, which is operated by the Association of Universities for Research in Astronomy, Inc., for NASA, under contract NAS5-26555.
G.M. acknowledges financial support from the Severo Ochoa grant  CEX2021-001131-S and from the Ramón y Cajal grant RYC2022-037854-I  funded by MCIN/AEI/1144 10.13039/501100011033 and FSE+.
J.M.A.M. is supported by the National Science Foundation (NSF) Graduate Research Fellowship Programme (GRFP) under Grant No. DGE-1842400. J.M.A.M. and N.M.B. acknowledge support from NASA'S Interdisciplinary Consortia for Astrobiology Research (NNH19ZDA001N-ICAR) under award number 19-ICAR19\_2-0041. The postdoctoral fellowship of K.B. is funded by F.R.S.-FNRS grant T.0109.20 and by the Francqui Foundation.
G.D. acknowledges the support of Magdalen College, Oxford. 
This publication benefits from the support of the French Community of Belgium in the context of the FRIA Doctoral Grant awarded to M.T.
Author F.J.P acknowledges financial support from the Severo Ochoa grant CEX2021-001131-S MICIU/AEI/10.13039/501100011033 and Ministerio de Ciencia e Innovación through the project PID2022-137241NB-C43

\end{acknowledgements}


\bibliographystyle{aa}
\bibliography{bibliography}

\begin{appendix} %First appendix
\onecolumn
\section{\tess\ target pixel files}\label{appendix:tpf}

\begin{figure*}[h!]
\centering
  \includegraphics[width=0.32\linewidth]{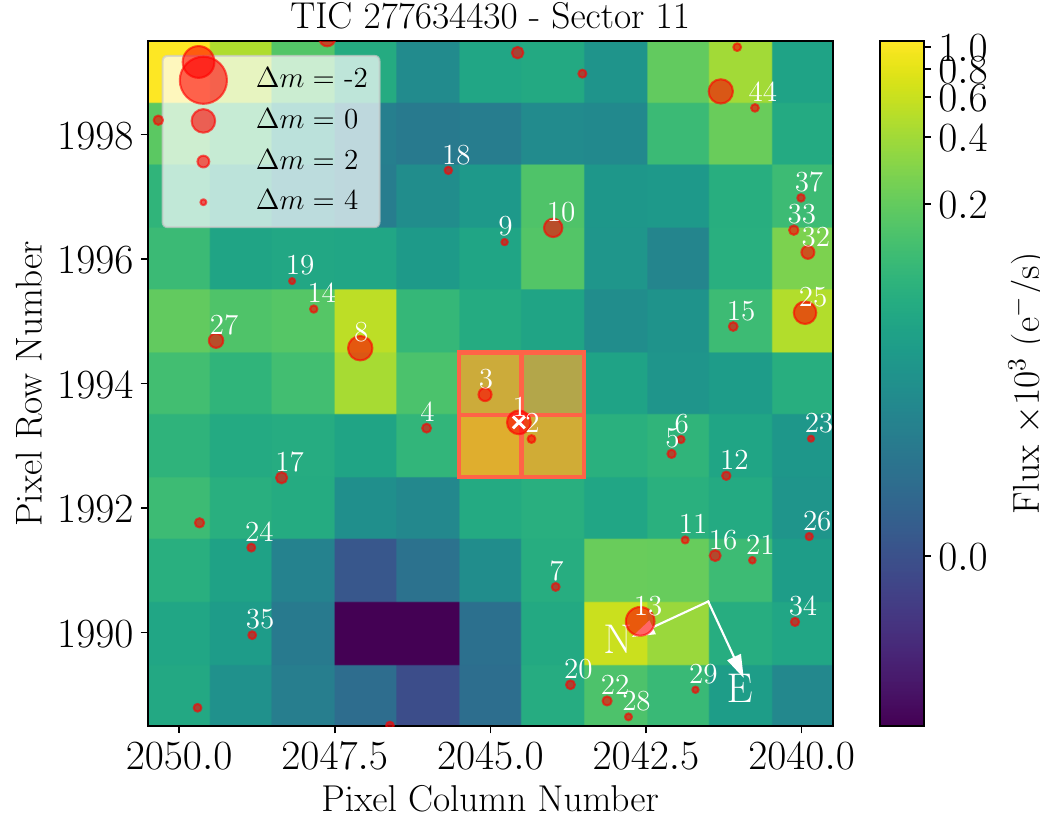}  
  \includegraphics[width=0.32\linewidth]{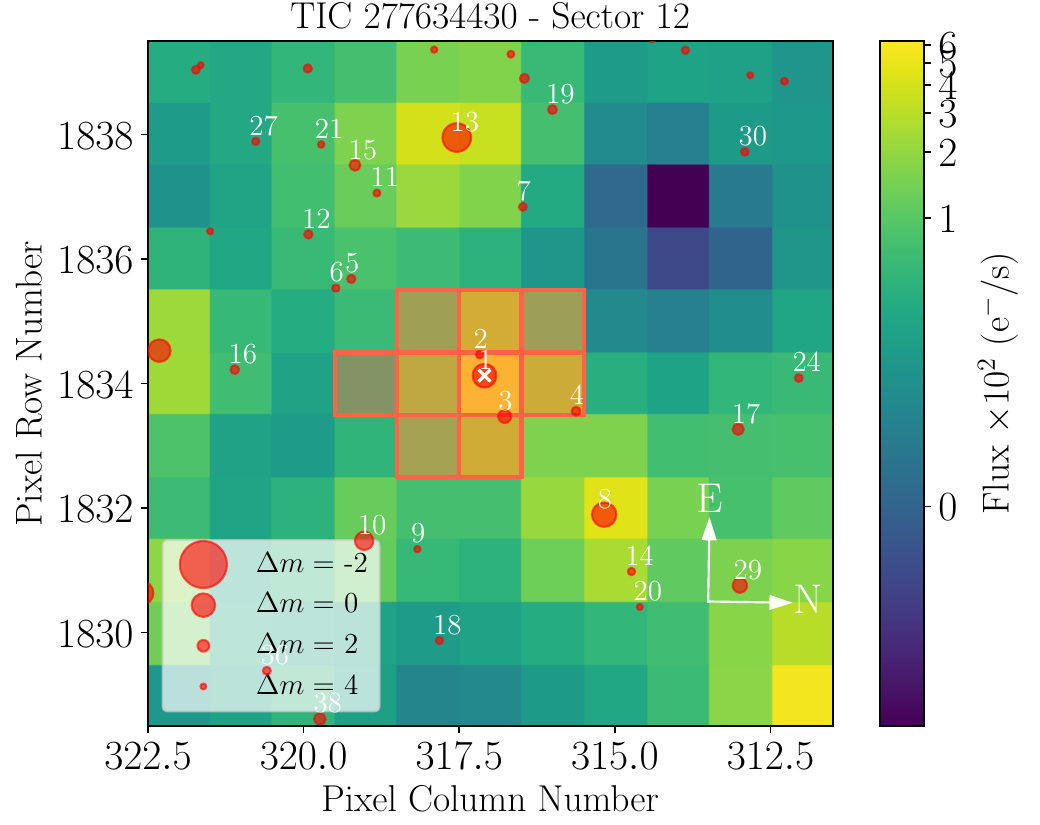}
  \includegraphics[width=0.32\linewidth]{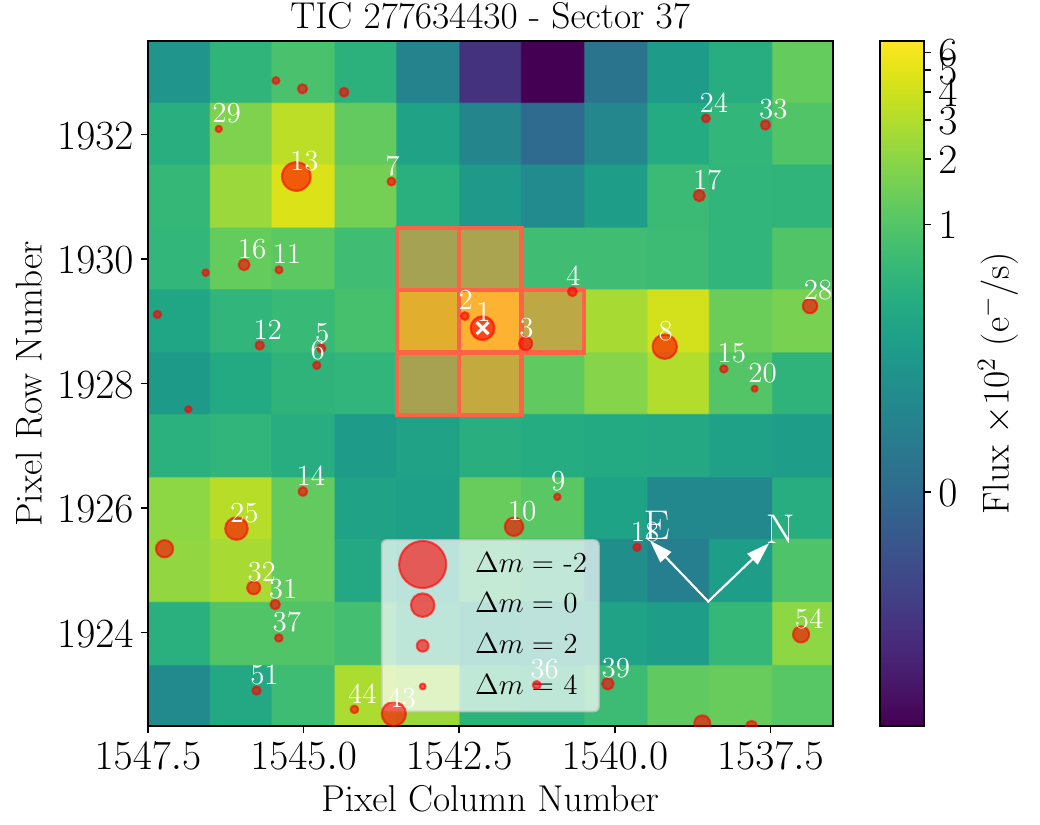}\\[1em]
  \includegraphics[width=0.32\linewidth]
  {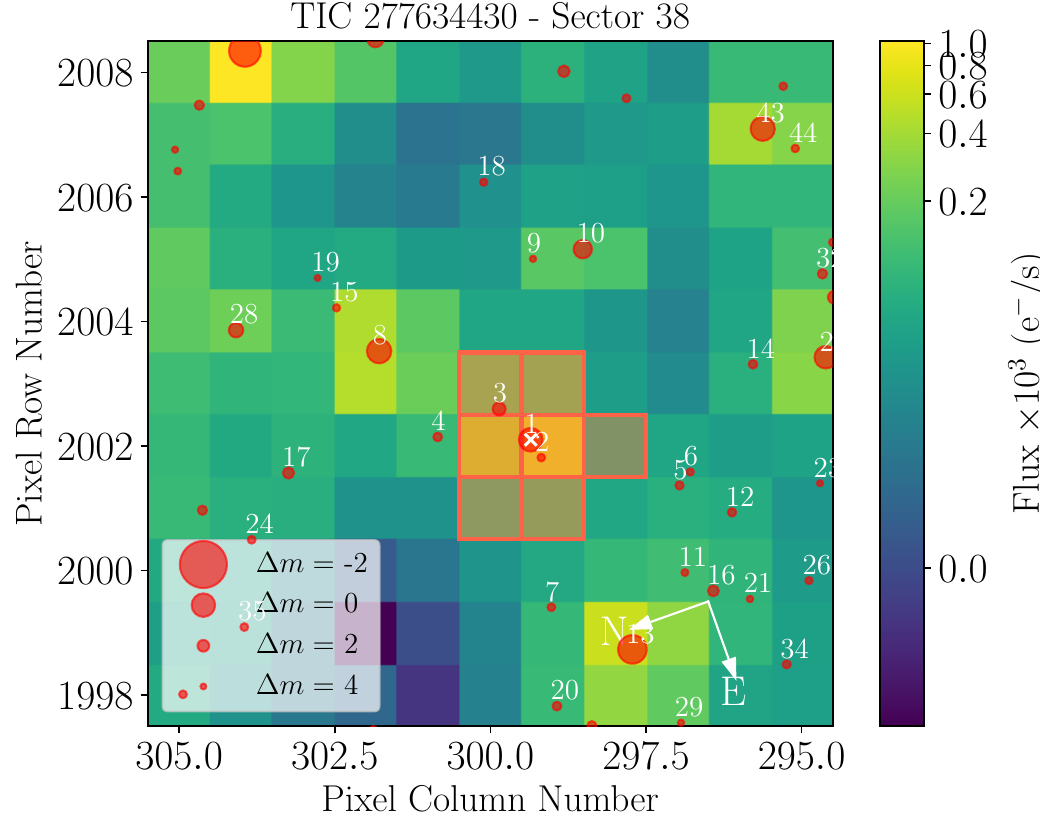}
  \includegraphics[width=0.32\linewidth]{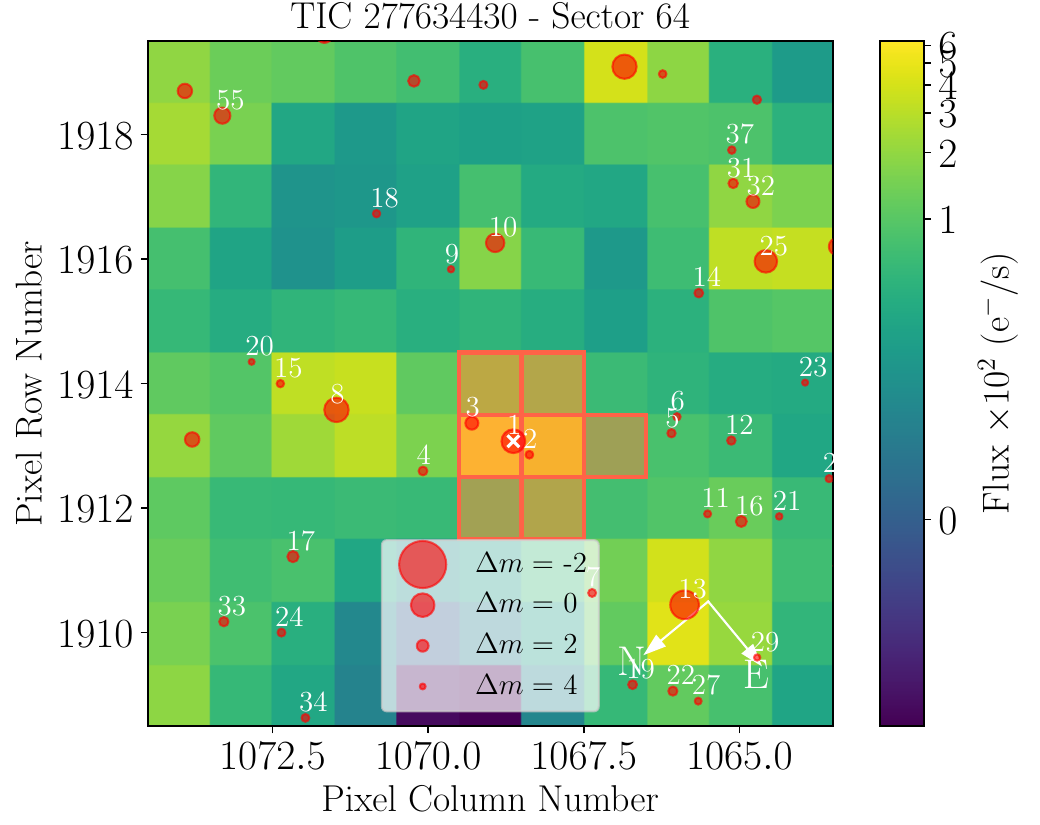}
  \includegraphics[width=0.32\linewidth]{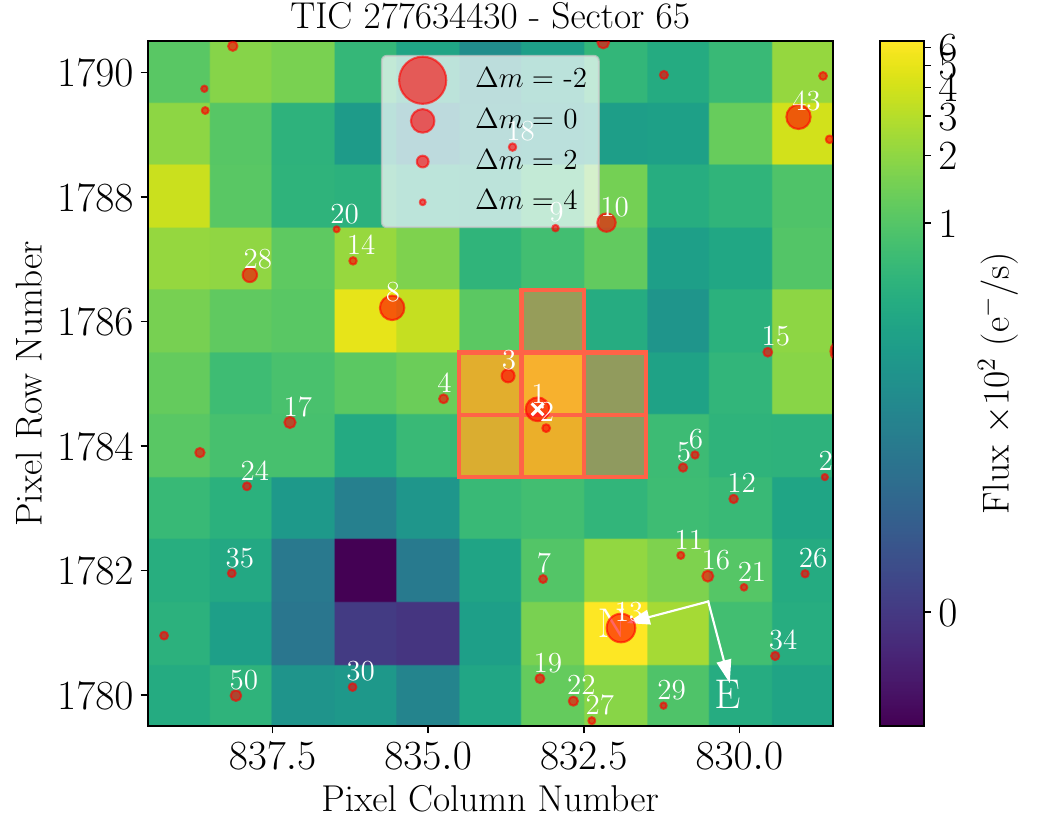}
  \caption{Same as Fig.~\ref{fig:tess_tpf}, but for {\it TESS} sectors 11, 12, 37, 38, 64, 65.
  }
    \label{fig:tess_tpf_all}
\end{figure*}

We report in Fig.~\ref{fig:tess_tpf_all} the TPF images of TOI-771 for all {\it TESS} sectors except for sector 10, which is shown in Fig.~\ref{fig:tess_tpf}.

\clearpage
\twocolumn
\section{{\it TESS} light curves}\label{appendix:light_curves}

\begin{figure}[h!]
\centering
  \includegraphics[width=\linewidth]{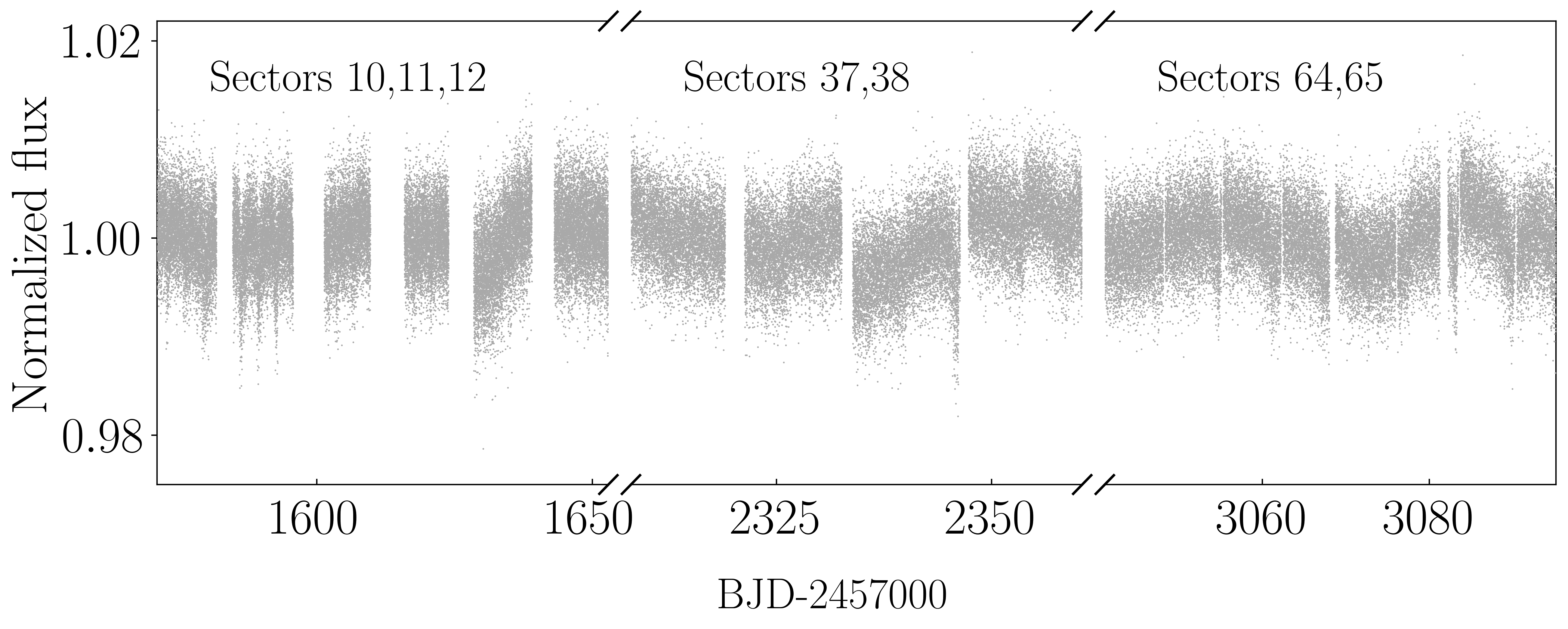}
  \includegraphics[width=\linewidth]{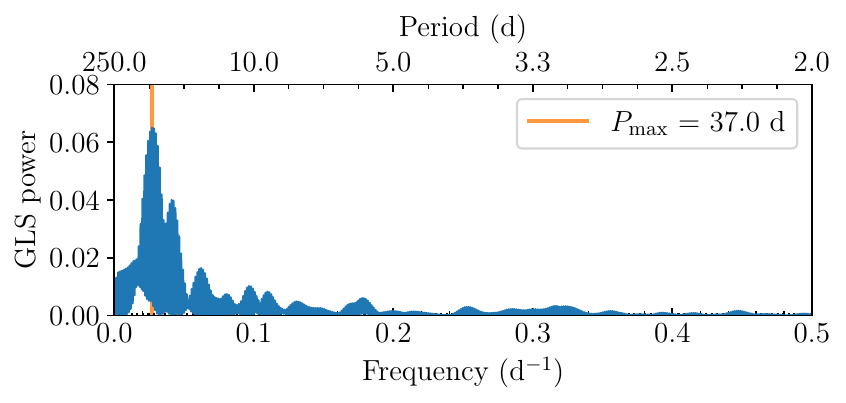}
  \caption{Top: TOI-771 {\it TESS}  SAP light curves of TOI-771, grouped in consecutive sectors. Bottom: GLS periodogram of the whole TESS SAP light curve, with the most significant peak marked with a vertical orange line. The dashed line shows the 1\% FAP probability.
  }
    \label{fig:tess_sap}
\end{figure}

We show in Fig.~\ref{fig:tess_sap} the TESS SAP light curves of TOI-771 and the related GLS periodogram, searching for the rotational period of the star at $\sim 98$~d (Sect.~\ref{sec:star_activity}). Such a long-period signal is difficult to identify in the current TESS photometry, as the clustering of the sectors in consecutive groups of two or three (S10, S11, S12; S37, S38; S64, S65), spread along the $\sim 4$ years of observations favour a shorter period of $\sim 37$~d, which is however not identified in any of the other activity indicators.

\begin{figure}[h!]
\centering
  \includegraphics[width=\linewidth]{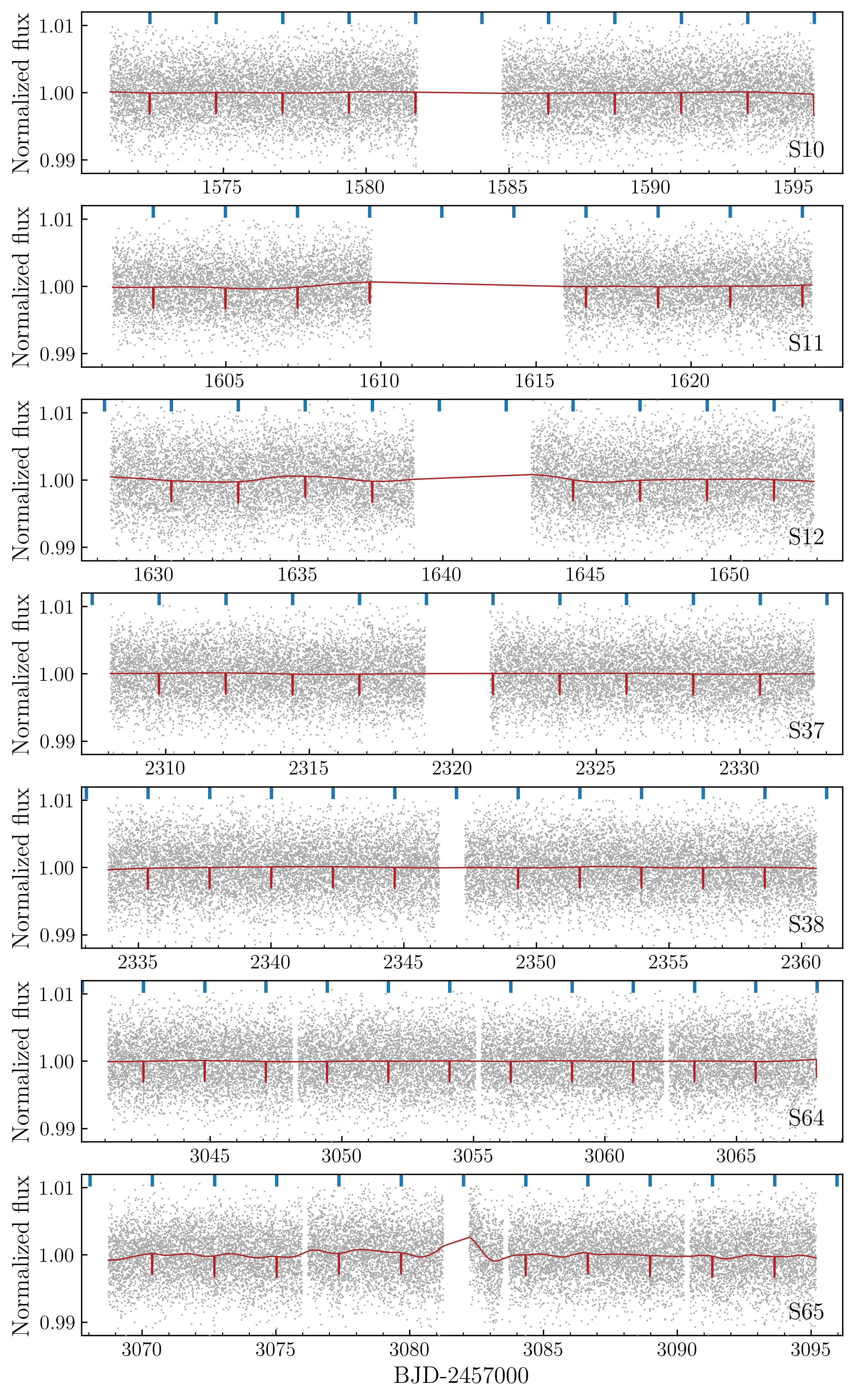}
  \caption{TOI-771 PDCSAP light curves of all available {\it TESS} sectors. The solid red line shows the photometric model including the Matérn-3/2 kernel GP regression  (Sect.~\ref{sec:photometric_fit_only}. The vertical blue lines highlight the transit ephemeris of TOI-771 b.
  }
    \label{fig:tess_lc}
\end{figure}

\begin{figure}[h!]
\centering
  \includegraphics[width=\linewidth]{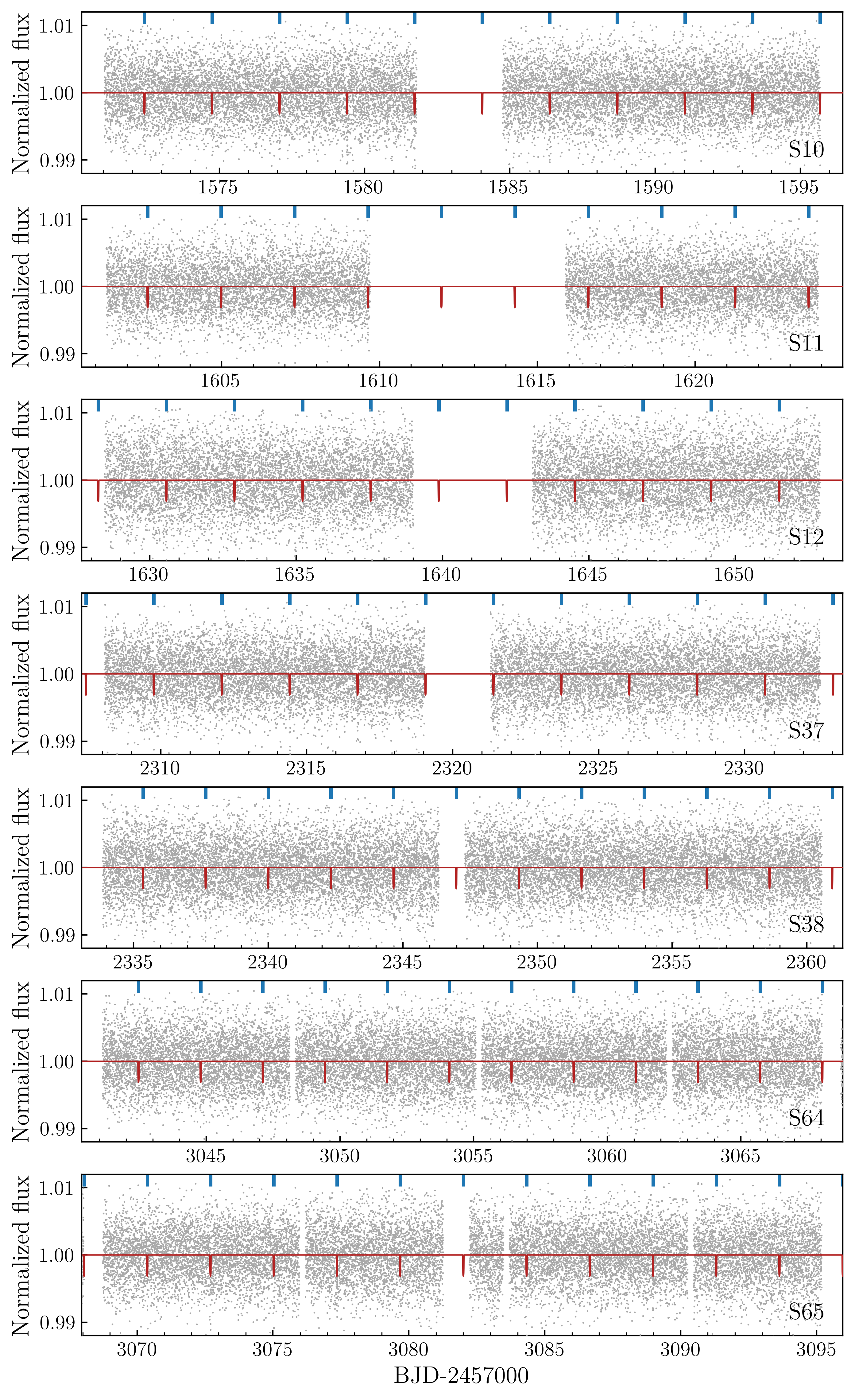}
  \caption{Same as Fig.~\ref{fig:tess_lc}, but using pre-detrended \textit{TESS} light curves with no GP regression model. The solid red line shows the best-fitting model (Sect.~\ref{sec:joint_fit}). 
  }
    \label{fig:tess_lc_detrended}
\end{figure}

We show in Figs.~\ref{fig:tess_lc} the \textit{TESS} light curves of TOI-771 modelled with the Matérn-3/2 kernel GP regression as described in Sect.~\ref{sec:photometric_fit_only}). 
Figure~~\ref{fig:tess_lc_detrended} shows the pre-detrended \textit{TESS} light curves with the best-fitting model as derived from the joint analysis described in Sect.~\ref{sec:joint_fit}.

\section{RV data}
We list in Table~\ref{table:RV_table_ESPRESSO} the ESPRESSO RVs and activity indicators collected for TOI-771.
\begin{table*}[h!]
\tiny
  \caption{ESPRESSO RV measurements and activity indicators of TOI-771.}
\label{table:RV_table_ESPRESSO}      % is used to refer this table in the text
\centering                                      % used for centering table
\resizebox{\linewidth}{!}{\begin{tabular}{c c c c c c c c c c}          % centered columns 
\hline\hline                        % inserts double horizontal lines
  BJD$_{\rm TDB}$ & RV & dLW  & CRX & Na \textsc{i} D$_1$ & Na \textsc{i} D$_2$ & H$\alpha$ & FWHM & BIS & $S$-index\\
 $($d$)$  & (\ms ) & (m$^2$ s$^{-2}$) & (\ms) & (dex)  & (dex) & (dex) & (\ms) & (\ms) & \\
\hline                                 % inserts single horizontal
2460284.824724  & $-4.39 \pm 0.50 $& $1.38 \pm  0.21 $&  $6.0 \pm  3.3 $  & $0.3312 \pm 0.0059 $ & $0.2534 \pm 0.0053 $  &   $0.9510 \pm 0.0028$ & $2716.2 \pm 2.6 $ & $-496.4 \pm 2.5 $ & $0.946 \pm 0.050$\\
2460289.774452  & $ -5.35 \pm 0.58$  & $1.10 \pm  0.28 $ & $7.6 \pm  3.9 $ &  $0.3620 \pm 0.0069 $   & $0.2665 \pm 0.0061$  &  $ 0.9289 \pm 0.0030$ & $2718.7 \pm 2.9 $& $-562.0 \pm 2.9 $ & $1.002 \pm 0.070$\\
2460290.812825  &  $1.84  \pm  0.59$ & $1.21 \pm  0.25 $ &  $-3.0 \pm  4.1 $ &  $0.3408 \pm 0.0067 $   & $0.2535 \pm 0.0060 $  &  $0.9657 \pm 0.0031 $ & $22719.8 \pm 2.9 $& $-264.5 \pm 2.9 $ & $1.172 \pm 0.064$\\
... & ... & ... & ... & ... & ... & ... & ... & ... & ...\\
\hline
\end{tabular}}
\tablefoot{This table is available in its entirety in machine-readable form.}
\end{table*}

\section{TTV search}\label{app:ttv}

\begin{figure}[h!]
\centering
  \includegraphics[width=\linewidth]{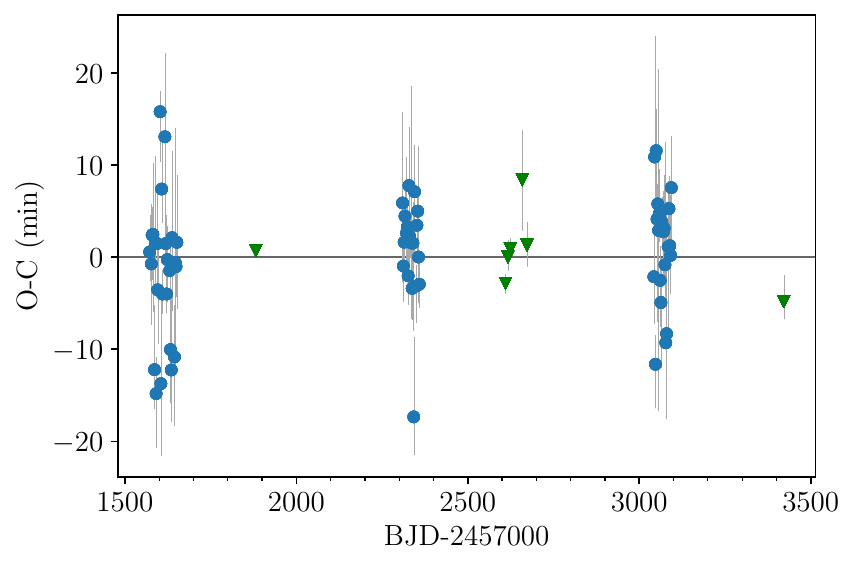}
  \caption{TTVs of TOI-771 b, measured computing the observed (O) minus calculated (C) transit times. Blue points show \textit{TESS} transits, while green triangles refer to ground-based transit observations.
  }
    \label{fig:TTV}
\end{figure}

We show in Figure~\ref{fig:TTV} the TTVs of TOI-771 b computed as described in Section~\ref{sec:joint_fit}.

\end{appendix}

\end{document}